\documentclass[twoside, 11pt]{article} 
\usepackage{epsfig,graphics} 
\usepackage{amssymb} 
\usepackage{amsmath} 
 
 \usepackage{latexsym,epsfig,bm,amssymb}
\usepackage{color}
\usepackage{amsthm,mathrsfs}

 \usepackage{mathptmx}

\newcommand\mC{{\mathbb C}} 
 \newcommand\R{{\mathbb R}}

\newcommand{\bce}{\begin{center}} 
\newcommand{\ece}{\end{center}}

\newcommand{\llr}{-\kern-.50em -\kern-.50em -\kern-.50em\longrightarrow}

\newcommand{\ds}{\displaystyle} 
\newcommand{\ov}{\overline} 
 
\newcommand{\pa}{\partial} 
\newcommand{\na}{\nabla} 
\newcommand{\rot}{{\rm rot~}} 
\renewcommand{\div}{{\rm div~}} 
\newcommand{\supp}{{\rm supp\5}}

 \newcommand{\ti}{\tilde} 
 \newcommand{\pr}{\prime} 
 \newcommand{\fr}{\frac} 
 
 \newcommand{\al}{\alpha} 
\newcommand{\si}{\sigma} 
\newcommand{\ve}{\varepsilon} 
\newcommand{\vp}{\varphi} 
\newcommand{\co}{{\rm const}} 
\newcommand{\ga}{\gamma} 
\newcommand{\de}{\delta} 
 
\newcommand{\om}{\omega} 
\newcommand{\Om}{\Omega} 
\newcommand{\De}{\Delta} 
 
 \newcommand{\ka}{\kappa} 
\newcommand{\lam}{\lambda}

\newcommand{\5}{{\hspace{0.5mm}}} 
 
\newcommand{\Si}{\Sigma}

\newcommand{\bA}{{\bf A}}

\newcommand{\cH}{{\cal H}} 
\newcommand{\cM}{{\cal M}} 
\newcommand{\cO}{{\cal O}} 
 
\newcommand{\cQ}{{\cal Q}} 
\newcommand{\cP}{{\cal P}}\newcommand{\cS}{{\cal S}} 
  \newcommand{\cT}{{\cal T}}
  \newcommand{\cX}{{\cal X}}
 \newcommand{\cZ}{{\cal Z}}
 
\newcommand{\cE}{{\cal E}}

\newcommand{\rRe}{{\rm Re\5\5}} 
\newcommand{\rIm}{{\rm Im\5\5}} 
 
\newcommand{\toAA} 
{\stackrel{\cal AA}{\mathop{\longrightarrow}}} 
 
\newcommand{\toQM} 
{\stackrel{\cal QM}{\mathop{\longrightarrow}}} 
 
\newcommand{\toL}{\stackrel{L^1}{\mathop{\longrightarrow}}}

\renewcommand{\theequation}{\thesection.\arabic{equation}} 
\newtheorem{theorem}{Theorem}[section] 
\renewcommand{\thetheorem}{\arabic{section}.\arabic{theorem}} 
\newtheorem{definition}[theorem]{Definition} 
 
\newtheorem{lemma}[theorem]{Lemma} 
 
\newtheorem{remark}[theorem]{Remark}

\newtheorem{pro}[theorem]{Proposition} 
\newcommand{\bob}{{\hfill\loota}} 
\newcommand{\loota} 
{\hbox{\enspace{\vrule height 10pt depth 0pt width 10pt}}}

\begin{document} 
 \bce 
{\huge\bf 
Attractors of nonlinear Hamilton PDEs} 
\\~
\\
{\large A.\,I.\,Komech}\footnote{
 The research was carried out at the IITP RAS at the expense of the Russian
Foundation for Sciences (project  14-50-00150).} 
\\ 
\textit{Faculty of Mathematics of Vienna University 
\\ 
Institute for Information Transmission Problems RAS} 
\\ 
alexander.komech@univie.ac.at 
\ece 
 
\abstract{ 
This is a~survey of  results on long time behavior and attractors for 
nonlinear Hamiltonian partial differential equations, considering the global 
attraction to stationary states, stationary orbits, and  solitons, 
the adiabatic effective dynamics of the solitons, and the asymptotic 
stability of 
the solitary manifolds.  The corresponding numerical results 
and  relations to quantum postulates are considered. 

This theory differs significantly from the theory of attractors of dissipative systems
where the attraction to stationary states is due to an energy dissipation caused by a friction.
For the Hamilton equations the friction and  energy dissipation are absent, and the 
attraction is caused by radiation which brings the energy irrevocably to infinity.

} 
 
 \vspace{-4mm}
\tableofcontents 
 
\section{Introduction} 
 
Our aim in this paper is to survey the results on  long time behavior and attractors for nonlinear Hamilton  
partial differential equations that appeared since 1990. 
 
 Theory of attractors for  nonlinear PDEs originated from the
 seminal paper of Landau \cite{L1944} published in 1944, where he
suggested  the first
 mathematical interpretation of turbulence as the growth of the
 dimension of attractors of  the Navier--Stokes equations when the
 Reynolds number increases.

The starting point for the corresponding mathematical  theory was 
provided 
in 1951 by Hopf who established for the first time the existence of
global solutions to  the 3D Navier--Stokes equations \cite{Hopf1951}.
He  introduced the `method of compactness' which is a nonlinear version 
of the Faedo-Galerkin approximations. This method relies on a priori 
estimates and Sobolev  embedding theorems. It has strongly influenced 
the development of the theory of nonlinear PDEs, see \cite{Lions1969}.

The modern development 
of the theory of attractors for general 
\textit{dissipative systems}, i.e. systems with friction 
(the Navier--\allowbreak Stokes equations, nonlinear parabolic equations, reaction-diffusion equations, 
wave equations with friction, etc.), as originated in the 1975--1985's in the works of Foias, Hale, 
Henry, Temam, and others \cite{FMRT2001, H1988, H1981}, 
was developed further in the works of Vishik, Babin, Chepyzhov, and others \cite{BV1992, CV2002}. 
A typical result of this theory in the absence of external excitation is the global convergence 
to a~steady state: 
for any finite energy solution, there is a convergence
\begin{equation}\label{at11} 
\psi(x,t) \to S(x), \qquad t\to + \infty 
\end{equation} 
in a~region $\Om\subset\mathbb R^n$ 
where $S(x)$ is a~steady-state solution with appropriate boundary conditions, and this
convergence holds as a~rule in the $L^2(\Om)$-metric. 
 In particular, the relaxation to an equilibrium regime 
in chemical reactions is followed by the energy dissipation. 
\medskip 
 
The development of a similar theory for the \textit{Hamiltonian PDEs} 
seemed unmotivated and impossible 
in view of energy conservation and time reversal for 
 these equations. However, as it turned out, such a~theory is possible and its shape was suggested 
 by a~novel mathematical interpretation of the fundamental postulates of quantum theory: 
\medskip\\ 
I. Transitions between quantum stationary orbits (Bohr 1913, \cite{Bohr1913}). 
\smallskip\\ 
II. The wave-particle duality (de Broglie 1924). 
\smallskip\\ 
\noindent Namely, postulate I can be interpreted 
as a~global attraction of all quantum 
trajectories to an attractor formed by stationary orbits, and II, as 
similar global attraction to solitons  \cite{K2013}. 
 
The investigations of the 1990--2014's showed that such long time asymptotics of solutions are in fact typical 
for a~number of nonlinear Hamiltonian PDEs. These results are presented in this article.  
This theory differs significantly from the theory of attractors of dissipative systems
where the attraction to stationary states is due to an energy dissipation caused by a friction.
For the Hamilton equations the friction and  energy dissipation are absent, and the 
attraction is caused by radiation which brings the energy irrevocably to infinity.
\medskip
 
The modern development of the theory of nonlinear Hamilton equations dates back to J\"orgens~\cite{Jor1961}, 
who has established
the existence of global solutions for 
nonlinear wave equations of the form 
\begin{equation}\label{Jw} 
\ddot\psi(x,t) = \De \psi (x,t) + F(\psi (x,t)), \qquad x \in \mathbb R^n, 
\end{equation}
developing the Hopf method of compactness. 
The subsequent studies were well reflected by J.-L. Lions in~\cite{Lions1969}.

First results on the long time asymptotics of solutions  were obtained by 
Segal \cite{Segal1966, Segal1968} and Morawetz and Strauss \cite{Mor1968, St68, MS1972}. In these papers 
the {\it local energy decay}  is proved for solutions to equations (\ref{Jw})
with {\it defocusing type} nonlinearities $F(\psi) = -m^2\psi-\kappa | \psi |^p \psi$, where 
$m^2\ge 0$, $\kappa>0$, and 
$p>1$. Namely, for sufficiently smooth and small initial states, one has
\begin{equation}\label{ledec}
\int_{|x|<R} [|\dot\psi(x,t)|^2+|\na\psi(x,t)|^2+|\psi(x,t)|^2]dx\to 0, \qquad t\to\pm\infty
\end{equation}
for any finite $R>0$. Moreover, 
the corresponding nonlinear wave and the scattering operators are constructed. 
In the works of Strauss \cite{St81-1, St81-2}, 
the completeness of scattering is established for small solutions to more general equations.

The existence of soliton solutions $\psi(x-vt)e^{i\om t}$ 
for a~broad class of nonlinear wave equations 
(\ref{Jw}) was extensively studied in the 1960--1980's. The most general results were obtained 
by Strauss, Berestycki and P.-L. Lions \cite{St77, BL83-1, BL83-2}. 
Moreover,
Esteban, Georgiev and S\'er\'e have constructed 
the solitons for the  nonlinear relativistically-invariant 
Maxwell--\allowbreak Dirac equations (\ref{DM}). 
The orbital stability of the solitons has been studied by Grillakis, Shatah, Strauss and others \cite{GSS87, GSS90}. 
 \medskip 
 
For convenience, the characteristic  properties of 
all finite energy solutions 
to an equation will be referred to as {\it global}, in order to distinguish them from the corresponding {\it local} 
properties for solutions with initial data sufficiently close to the attractor.

All the above-mentioned results \cite{Segal1966}--\cite{St81-2} 
on the  local energy decay (\ref{ledec}) mean that the 
corresponding local attractor of small initial states consists of the zero point only. 
First results on the global attractors for 
nonlinear Hamiltonian PDEs were obtained by the author 
in the 1991--1995's for 1D  models \cite{K1991, K1995a, K1995b}, and were 
later extended to nD equations. 
The main difficulty here is due to the absence of energy dissipation for the Hamilton equations. 
For example, the attraction to a~(proper) attractor is impossible for any finite-dimensional 
Hamilton system because of the energy conservation. 
The problem is attacked by analyzing the energy radiation to infinity, which plays the role of dissipation. 
The progress  relies on a~novel application 
of subtle methods of harmonic analysis:  the Wiener Tauberian theorem, 
the Titchmarsh convolution theorem, theory of quasi-measures, 
the Paley-Wiener estimates, the eigenfunction expansions for nonselfadjoint Hamilton operators based on 
M.G.~Krein theory of $J$-selfadjoint operators,
and others. 
 
The results obtained so far indicate a~certain dependence 
of long-teme asymptotics of solutions on the symmetry group of the equation: 
for example, it may be the trivial group 
$G = \{e\} $, or the unitary group $G = U(1) $, or the group of translations 
 $G = \mathbb R^n$.  
 Namely, 
 the corresponding  results suggest that 
 for `generic' autonomous 
 equations with a Lie symmetry  group $G$,
 any finite energy solution  admits the asymptotics 
\begin{equation}\label{at10} 
\psi(x,t) \sim e^{g_\pm t} \psi_\pm (x) , \qquad t \to \pm \infty. 
\end{equation} 
Here, $e^{g_\pm t}$ is 
a representation of 
the one-parameter subgroup of the symmetry group $G$ 
which corresponds to the generators $g_\pm$ from the corresponding Lie algebra, 
while $\psi_\pm(x) $ are some 
`scattering states' depending on the considered trajectory $\psi(x,t)$,
with each pair $(g_\pm, \psi_\pm)$ being a~solution to the corresponding nonlinear eigenfunction problem. 
\smallskip 
 
For the trivial symmetry group 
 $G =\{e\}$, the conjecture (\ref{at10}) means the global attraction to the corresponding steady states 
\begin{equation}\label{ate} 
\psi(x,t) \to S_\pm(x) , \qquad t\to\pm\infty 
\end{equation} 
(see Fig.~\ref{fig-1}).
Here $S_\pm(x) $ are some stationary 
states depending on the considered trajectory $\psi(x,t) $, 
and 
the convergence holds in local seminorms of type $L^2(|x|<R)$ for
 any $R>0$. 
The convergence (\ref{ate}) in global  norms (i.e., corresponding to 
$R=\infty$)  cannot hold  due to the
energy conservation. 
 
 In particular,
the asymptotics (\ref{ate})
can be easily demonstrated for  
the d'Alembert equation, see (\ref{dal})-- (\ref{dal3}). 
In this example 
the convergence (\ref{ate}) in global  norms 
obviously fails due to presence of travelling waves $f(x\pm t)$.
\smallskip

Similarly, for the unitary symmetry group $G=U(1)$, the asymptotics (\ref{at10}) means  the global attraction 
to `stationary orbits' 
\begin{equation}\label{atU} 
\psi(x,t)\sim\psi_\pm(x) e^{-i \om_\pm t} , \qquad t \to \pm \infty 
\end{equation} 
in the same local seminorms  (see Fig.~\ref{fig-3}). 
These asymptotics were inspired  by Bohr's postulate on transitions 
between quantum stationary states (see Appendix for details).

Our results confirm such asymptotics 
for generic $U(1)$-invariant  nonlinear  
equations of type (\ref{KG1}) and (\ref{KGN})--(\ref{Dn}).
More precisely, we have proved the global attraction {\it to the manifold
of the stationary orbits}, though 
the attraction to the concrete stationary orbits, with fixed $\om_\pm$, is still open problem.

Let us emphasize that we conjecture
the asymptotics (\ref{atU}) for {\it generic} $U(1)$-invariant equations.
This means that   the long time behavior
may be quite different for 
$U(1)$-invariant
equations of `positive codimension'. In particular, 
for linear Schr\"odinger equation
\begin{equation}\label{lS} 
i\dot\psi(x,t)=-\De\psi(x,t)+V(x)\psi(x,t),\qquad x\in\R^n
\end{equation} 
the asymptotics (\ref{atU}) generally fail. Namely, any finite energy solution 
admits the spectral representation 
\begin{equation}\label{lSs} 
\psi(x,t)=\sum C_k\psi_k(x)e^{-i\om_k t}+\int_0^\infty C(\om)\psi(\om,x)e^{-i\om t}d\om,
\end{equation}
where $\psi_k$ and $\psi(\om,\cdot)$ are the corresponding eigenfunctions of the discrete and continuous 
spectrum, respectively. The last integral is a dispersion wave which decays to zero in  
local seminorms $L^2(|x|<R)$ for any $R>0$ (under appropriate conditions on the potential $V(x)$).
Respectively, the attractor is the linear span of the eigenfunctions $\psi_k$.
However, the long-time 
asymptotics  does not reduce to a single term like (\ref{atU}),  so  the linear case is degenerate
in this sense. 
Let us note that our results for equations  (\ref{KG1}) and (\ref{KGN})--(\ref{Dn}) are established 
for {\it strictly nonlinear case}: see the condition 
(\ref{C4}) below, which eliminates linear equations. 
\smallskip

Finally, for the symmetry group of translations 
 $G =\mathbb R^n$, the asymptotics  (\ref{at10}) means the global attraction  
to solitons (traveling wave solutions)
\begin{equation}\label{att} 
\psi(x,t) \sim \psi_\pm (x-v_\pm t) ,\qquad t\to \pm \infty, 
\end{equation} 
for  generic translation-invariant equation. In this case we conjecture
that the 
the convergence holds in the local seminorms 
in the comoving frame, i.e., in
$L^2 (|x-v_\pm t|<R)$ for any $R>0$. 
In particular, $\psi(x,t)=f(x-t)+g(x+t)$ for any solution to the 
d'Alembert equation (\ref{dal}). 
\smallskip 

For more sophisticated symmetry groups $G=U(N)$, the asymptotics
(\ref{at10}) means the attraction to $N$-frequency trajectories, which
can be
quasi-periodic.  The symmetry groups $SU(2)$, $SU(3)$ and others
were suggested in 1961  by 
Gell-Mann and Ne'eman for the strong interaction of  baryons \cite{GM1962, Ne1962}. 
The suggestion relies on the discovered parallelism between empirical data
for the baryons,
and the `Dynkin scheme'
of Lie algebra $su(3)$ with $8$ generators (the famous `eightfold way'). 
This theory resulted in
the scheme of quarks and in the
development of the
quantum  chromodynamics \cite{AF1963,HM1984},
and  in the prediction of a new baryon 
with prescribed values of 
its  mass and decay products. This particle, the $\Om^-$-hyperon,
was promptly discovered experimentally  \cite{omega-1964}.

This empirical correspondence between the Lie algebra generators and elementary
particles 
presumably gives an
evidence in favor of the general conjecture
(\ref{at10}) for equations with the Lie symmetry groups. 
\medskip
 
Let us note that our conjecture (\ref{at10})  specifies the concept of 
"localized solution/coherent structures"
from 
"Grande Conjecture" and "Petite Conjecture"  of Soffer \cite[p.460]{soffer2006}
in the context of $G$-invariant equations.
The  Grande Conjecture is proved 
in \cite{KM2009a}
for 1D wave equation coupled to a nonlinear oscillator (\ref{w1})\; see Theorem \ref{t2}.
Moreover, a suitable version of the  Grande Conjecture is also proved 
in
\cite{IKS2004a}--\cite{IKS2004b}  
for 3D wave, Klein-Gordon and Maxwell equations coupled to a relativistic particle  
with sufficiently small charge
\eqref{rosm};
see  Remark \ref{rGC}.
Finally, for any matrix  symmetry group $G$,
(\ref{at10}) 
implies the 
Petite Conjecture 
since 
the localized solutions $e^{g_\pm t}\psi_\pm(x)$ 
are quasiperiodic then.

 \smallskip
 
Now let us dwell upon the available results on the asymptotics \eqref{ate}--\eqref{att}. 
\smallskip\\ 
\textbf{I. Global attraction to stationary states (\ref{ate})} was first established by the author in \cite{K1991}--\cite{K2000} 
for the one-dimensional 
wave equation coupled to nonlinear oscillators (equations (\ref{w1}), (\ref{wN})) and for equations with 
general space-localized
nonlinearities (equation (\ref{neli})). 
 
These results were extended by the author in collaboration with Spohn and Kunze in \cite{KSK1997, KS2000} 
to the three-dimensional wave equation coupled to a particle \eqref{w3}--\eqref{q3}
under the Wiener condition \eqref{W1} on the charge density of the particle, 
and to the similar 
Maxwell-Lorentz equations \eqref{ML}  (see the survey \cite{S2004}). 

In \cite{KM2009a}--\cite{KM2013}, the asymptotic completeness of scattering for nonlinear wave equation (\ref{w1})
was proved in collaboration with Merzon. 
 
These results rely on a~detailed study of energy radiation to infinity. 
In \cite{K1991}--\cite{K1995b} and 
\cite{KM2009a}--\cite{KM2013} we justify this radiation by 
the `reduced equation'  \eqref{rede}, 
containing radiation friction and incoming waves, 
and in \cite{KSK1997, KS2000}, by a~novel integral representation for the radiated energy 
as the~convolution (\ref{Rsv}) and the application of the Wiener Tauberian theorem. 
\smallskip\\
\textbf{II. Local attraction to stationary orbits (\ref{atU})} (i.e., for initial states close to the set of stationary orbits) 
was first established by Soffer and Weinstein,   
Tsai and Yau,  and others 
 for nonlinear Schr\"odinger, wave 
and Klein--\allowbreak Gordon equations with external potentials  
under various types of spectral assumptions on the linearized dynamics \cite{W1985}--\cite{T2003}. 
However, no examples of nonlinear equations with the desired spectral properties were constructed. 
Concrete examples have been constructed by the author together with Buslaev, Kopylova and Stuart 
in \cite{BKKS2008, KKopSt2011} 
for one-dimensional Schr\"odinger equations coupled to nonlinear oscillators. 
 
The main difficulty of the problem is that the soliton dynamics is unstable along the solitary manifold, since the 
distance between solitons with arbitrarily close velocities increases indefinitely in time. 
However, the dynamics can be stable in the transversal symplectic-orthogonal directions 
to this manifold.
\medskip\\
\textbf{Global attraction to stationary orbits (\ref{atU})} was obtained for the first time by the author in \cite{K2003} 
for  the one-dimensional 
Klein--\allowbreak Gordon equation coupled to a~$U(1)$-invariant oscillator (equation \eqref{KG1}). 
The proofs rely on a~novel analysis of the energy radiation with the 
application of quasi-measures and the Titchmarsh convolution theorem (Section~3). 
These results and methods were further developed by the author in collaboration with A.\,A.~Komech \cite{KK2006, KK2007}, 
and were extended in \cite{KK2010a, KK2008} 
to a~finite number of $U(1)$-invariant oscillators (equation (\ref{KGN})), and 
in \cite{KK2009, KK2010b} to the $n$-dimensional Klein--\allowbreak Gordon and Dirac equations 
coupled to $U(1)$-invariant oscillators via a nonlocal interaction (equations (\ref{KGn}) and (\ref{Dn})). 
 
Recently, the global attraction to stationary orbits was established for discrete in space and time 
nonlinear Hamilton equations \cite{C2013}. The proofs required a~refined version of the Titchmarsh convolution theorem 
for distributions on the circle \cite{KK2013}.

The main ideas of the proofs \cite{K2003}--\cite{C2013} rely on the 
radiation mechanism caused by dispersion radiation and nonlinear inflation of spectrum  (Section~\ref{Sec-4.8}). 
\smallskip\\
\textbf{III. Attraction to solitons} was first discovered in 1965 by Zabusky and Kruskal  
in numerical simulations of the Korteweg--de Vries equation (KdV). Subsequently, 
global asymptotics of the type 
\begin{equation}\label{at14} 
\psi(x,t)\sim \sum\psi^k_\pm(x-v^k_\pm t)+w_\pm(x,t),\qquad t \to\pm\infty, 
\end{equation} 
were proved for finite energy solutions to 
\textbf{integrable} Hamilton 
translation-invariant equations (KdV and others) 
by Ablowitz, Segur, Eckhaus, van Harten, and others (see \cite{EvH}). 
Here, each soliton $\psi^k_\pm(x-v^k_\pm t) $ is a trajectory of the translation group $G=\mathbb R$, 
while $w_\pm(x,t)$ are some dispersion waves, and the asymptotics hold in a global norm like $L^2(\mathbb R)$.
\smallskip 
 
First results on the \textbf{local attraction to solitons for non-integrable equations} 
were established by Buslaev and Perelman  
for one-dimensional nonlinear 
translation-invariant Schr\"odinger equations in \cite{BP1993, BP1995}:
the strategy 
relies on symplectic projection
onto the solitary manifold
in the Hilbert phase space 
(see Section 6.2). 
 The key role of the symplectic structure is 
 explained by 
 the conservation of the symplectic form
by  the Hamilton dynamics. 
This strategy was completely justified in~\cite{BS2003}, thereby extending 
quite far the Lyapunov stability theory. 
The extension of this strategy to
the multidimensional translation-invariant Schr\"odinger equation was done by Cuccagna~\cite{C2001}

Further, for generalized KdV equation and the regularized long-wave equation, the local attraction to the solitons was 
established by Weinstein, Miller and Pego \cite{PW1994, MW1996}. 
Martel and Merle have extended these results to the subcritical gKdV equations \cite{MM2005}, and Lindblad and Tao 
have done this in the context of 1D nonlinear wave equations \cite{LT2012}. 
 
The general strategy \cite{BP1993}--\cite{BS2003} was developed in \cite{IKV2006}--\cite{IKV2011}  
for the proof of local attraction to solitons for the system of 
a~classical particle coupled to the Klein--\allowbreak Gordon, Schr\"odinger, Dirac, wave and Maxwell fields 
 (see the survey \cite{Im2013}). 
 \smallskip

 For relativistically-invariant equations 
  the first results on the local attraction to the solitons 
 were obtained by Kopylova and the author in the context of the nonlinear 
 Ginzburg--\allowbreak Landau equations \cite{K2002}--\cite{Kumn2013}, and 
 by  Boussaid and Cuccagna, for the nonlinear Dirac equations \cite{BC2012}. 
 \smallskip 
  
 In a series of papers, Egli, Fr\"ohlich, Gang, Sigal, and Soffer  have established the
 convergence to a~soliton with subsonic speed for a~tracer particle 
 with initial supersonic speed
 in the Schr\"odinger field.
 The convergence 
is considered as a model of the Cherenkov radiation, see \cite{FG2014} and the references therein. 
  \smallskip 
  
 The asymptotic stability of $N$-soliton solutions was studied by Martel, Merle and Tsai \cite{MMT2002}, 
 Perelman \cite{P2004}, and Rodnianski, Schlag and Soffer \cite{RSS2003,RSS2005}. 
 \smallskip 
 
One of the essential components of many works on local attraction to stationary orbits and solitons is the 
dispersion decay for the corresponding linearized Hamilton equations. The theory of this decay was developed 
by Agmon, Jensen and Kato for the Schr\"odinger equations \cite{Agmon,JK}, 
and was extended by the author and Kopylova to the wave and Klein--\allowbreak Gordon equations \cite{KopK2012-1}--\cite{KopK2010-10} 
(see also \cite{KKK2006}--\cite{K2010} for 
the discrete Schr\"odinger and Klein--\allowbreak Gordon equations). 
 \smallskip\\ 
\textbf{Global attraction to solitons (\ref{att}) for non-integrable equations} 
was established for the first time by the author together with 
Spohn \cite{KS1998} for a scalar wave field coupled to a relativistic particle (the system \eqref{wq3}) 
under the Wiener condition \eqref{W1} on the particle charge density. 
 In  \cite{IKM2004}, this result was extended 
 by the author in collaboration with Imaykin and Mauser 
 to a~similar Maxwell-Lorentz system with zero external fields \eqref{ML}. 
 The global attraction to solitons 
  was proved also   
  for a~relativistic particle 
  with sufficiently small charge
  in 3D wave, Klein--\allowbreak Gordon and Maxwell fields
  \cite{IKS2004a}--\cite{IKS2004b}. 
 
 These results give the first 
 rigorous justification of the {\it radiation damping} in classical electrodynamics 
 suggested by Abraham and Lorentz \cite{A1902, A1905}, see the survey \cite{S2004}.
 \smallskip 
 
For  relativistically-invariant  one-dimensional nonlinear wave equations (\ref{Jw})
global soliton asymptotics (\ref{at14}) 
were
confirmed by numerical simulations by Vinnichenko 
(see \cite{KMV2004} and also Section 7). 
However, the proof in the relativistically-invariant case remains an open problem. 
\medskip\\
\textbf{Adiabatic effective dynamics of solitons} means the evolution of states which are 
close to a~soliton with parameters depending on time (velocity, position, etc.) 
\vspace{-2mm}
\begin{equation}\label{asoli} 
\psi(x,t) \sim \psi_{v(t)} (x-q(t)). 
\end{equation} 
These asymptotics are typical for approximately translation-invariant 
systems with  initial states sufficiently close to the solitary manifold. 
 Moreover, in some cases it  turns out possible to find an `effective dynamics' describing the evolution of soliton parameters. 
 
Such adiabatic effective soliton dynamics was justified for the first time by the author 
together with Kunze and Spohn \cite{KKS1999} for a~relativistic particle coupled to 
a scalar wave field and a~slowly varying external potential  
(the system \eqref{w3}--\eqref{q3}). In \cite{KS2000ad}, 
this result was extended 
by Kunze and Spohn  
to a~relativistic particle coupled to the Maxwell field and to small external fields (the system \eqref{ML}). Further, Fr\"ohlich together with 
Tsai and Yau obtained similar results 
for nonlinear Hartree equations \cite{FTY2002}, and with Gustafson, Jonsson and Sigal, for nonlinear Schr\"odinger equations \cite{FGJS2004}. 
Stuart, Demulini and Long have proved similar results 
for nonlinear 
Einstein--\allowbreak Dirac, Chern--Simons--Schr\"odinger and Klein--\allowbreak Gordon--\allowbreak Maxwell systems \cite{S2010}--\cite{LS2009}. 
Recently, Bach, Chen, Faupin, Fr\"ohlich and Sigal 
proved the adiabatic effective dynamics for one electron in second-quantized Maxwell field 
in the presence of a~slowly varying external potential \cite{BCFJS}. 
 
\medskip 
 
Note that the attraction to stationary states \eqref{ate} resembles asymptotics 
of type (\ref{at11}) for dissipative systems. However, there are a~number of significant differences: 
\medskip\\ 
I. In the dissipative systems, attraction (\ref{at11}) is due to the energy dissipation. This attraction holds 
\\ 
$\bullet$ only as $t\to + \infty $; 
\\ 
$\bullet $ in bounded and unbounded domains; 
\\ 
$\bullet$ in `global' norms. 
\\ 
Furthermore, the attraction (\ref{at11}) holds for all solutions of finite-dimensional dissipative systems. 
\medskip\\
II. In the Hamilton systems, attraction \eqref{ate} is due to the energy radiation. This attraction holds 
\\ 
$\bullet$ as $t\to \pm \infty $; 
\\ 
$\bullet$ only in unbounded domains; 
\\ 
$\bullet$ only in local seminorms. 
\\ 
However, the attraction \eqref{ate} cannot hold for all solutions 
of any 
finite-dimensional Hamilton system 
with nonconstant Hamilton functional. 
\medskip 
 
In conclusion it is worth 
mentioning
that the analogue of asymptotics 
\eqref{ate}--\eqref{att} are not yet shown to hold for 
the fundamental equations of quantum physics (systems of the 
Schr\"odinger, Maxwell, Dirac, Yang--Mills equations and their second-quantized versions \cite{Sakurai}). 
The perturbation theory is of no avail here, 
since the convergence \eqref{ate}--\eqref{att} cannot be uniform on an 
infinite time interval. These problems remain open, and 
their analysis agrees 
with the Hilbert's sixth problem on the `axiomatization of theoretical physics', 
as well as with the spirit of Heisenberg's program for 
nonlinear theory of elementary particles \cite{Heis1961, Heis1966}. 
 
However, the main motivation for such investigations is to 
clarify dynamic description of fundamental quantum phenomena which play the key role 
throughout modern physics and technology: 
 the thermal and electrical conductivity of solids, the 
laser and synchrotron radiation, the photoelectric effect, the thermionic emission, the Hall effect, etc. 
The basic physical principles of these phenomena are already established, but their 
dynamic description 
as inherent properties of
fundamental equations still remains missing \cite{BLR2000}. 
\bigskip 
 
In Sections 2--4 we review the results on  global attraction 
 to a~finite-dimensional attractor consisting of stationary states, 
stationary orbits and solitons. 
In Section~5, we state the results on the adiabatic effective dynamics of solitons, 
and in Section~6, the results on the asymptotic stability of solitary waves. 
Section~7 is concerned with numerical simulation of soliton asymptotics for relativistically-invariant nonlinear wave equations. 
In Appendix~A we discuss the relation of global attractors to quantum postulates. 
\bigskip\\
\textbf{Acknowledgments.} 
I wish to express my deep gratitude to H.~Spohn  and B.~Vainberg for long-time
collaboration on attractors of
  Hamiltonian PDEs, as well as to A.~Shnirelman for
many  useful long-term discussions.  I~am  also grateful to V.~Imaykin,
A.\,A.~Komech, E.~Kopylova, M.~Kunze, A.~Merzon and D.~Stuart for collaboration
lasting many years. My special thanks go to E.~Kopylova for checking
the
manuscript and for numerous suggestions.

\setcounter{equation}{0} 
\section{Global attraction to stationary states} 
 
Here we describe the results on asymptotics \eqref{ate} with a nonsingleton attractor, which were obtained 
in the 1991--1999's for the Hamilton nonlinear PDEs. 
First results of this type  
were obtained for one-dimensional 
wave equations coupled to nonlinear oscillators \cite{K1991}--\cite{K2000}, 
and were later extended to the three-dimensional wave equation and Maxwell's equations coupled 
to relativistic particle \cite{KSK1997, KS2000}. 

 The global attraction \eqref{ate} can be easily demonstrated on the 
 trivial (but instructive) example of the d'Alembert equation:
 \begin{equation}\label{dal}
 \ddot\psi(x,t)=\psi''(x,t),\quad\psi(x,0)=\psi_0(x),\quad\dot\psi(x,0)=\pi_0(x), \qquad x \in \mathbb R.
 \end{equation} 
Let us assume that $\psi_0'(x)\in L^2(\R)$ and $\pi_0(x)\in L^2(\R)$, and moreover,
\begin{equation}\label{lim} 
\psi_0(x)
\xrightarrow
[x\to\pm\infty]{}
  C_\pm,\qquad \int_{-\infty}^{\infty}|\pi_0(x)|dx<\infty. 
\end{equation} 
 Then the d'Alembert formula gives
 \begin{equation}\label{dal2} 
 \psi(x,t)=\frac{\psi_0(x+t)+\psi_0(x-t)}2+\fr12\int_{x-t}^{x+t}\pi_0(y)dy
 \xrightarrow
[t\to\pm\infty]{}
 S_\pm(x)=
 \frac{C_+ +C_-}2\pm \fr12\int_{-\infty}^{\infty}\pi_0(y)dy
 \end{equation} 
 where the convergence holds uniformly on each finite interval $|x|<R$. Moreover, 
 \begin{equation}\label{dal3} 
 \dot\psi(x,t)=\frac{\psi_0'(x+t)-\psi_0'(x-t)}2+\frac{\pi_0(x+t)+\pi_0(x-t)}2
 \xrightarrow
[t\to\pm\infty]{} 0,
 \end{equation} 
 where the convergence holds in $L^2(-R,R)$ for each $R>0$. Thus, the attractor is the set 
 of $(\psi(x),\pi(x))=(C,0)$ where $C$ is any constant. Let us note that the limits (\ref{dal2}) generally are  
 different for positive and negative times.

\subsection{Lamb system: a string coupled to nonlinear oscillators} 
In \cite{K1991, K1995a}, asymptotics \eqref{ate} 
was obtained for the wave equation coupled to nonlinear oscillator 
\begin{equation}\label{w1} 
\ddot\psi(x,t)=\psi''(x,t) + \delta(x) F(\psi(0, t)), \qquad x \in \mathbb R. 
\end{equation} 
All the derivatives here and below are understood in the sense of distributions. 
Solutions can be scalars-valued or vector-valued, $\psi\in\mathbb R^N$. 
Physically, this is a~string in $\mathbb R^{N +1}$, coupled to an 
oscillator at $x = 0$ 
acting on the string with force $F(\psi(0, t))$ orthogonal to the string. 
For linear function $F(\psi)=-k\psi $, such a~system was first considered by H.~Lamb \cite{Lamb1900}. 
 
\begin{definition} 
{\rm $\cE$ denotes the Hilbert phase space of functions $(\psi(x),\pi(x))$ with finite norm 
\begin{equation}\label{cE} 
\Vert(\psi,\pi)\Vert_{\cE}=\Vert\psi'\Vert+|\psi(0)|+\Vert\pi\Vert, 
\end{equation} 
where $\Vert\cdot\Vert$ stands for the norm in $L^2:=L^2(\mathbb R)$.} 
\end{definition} 
 
We assume that the nonlinear force $F(\psi)$ is a~potential field; i.e., 
for a real function $U(\psi)$
\begin{equation}\label{FU} 
F(\psi)=-\na U(\psi), \qquad \psi\in\mathbb R^N; \qquad U(\psi)\in C^2 (\mathbb R^N).
\end{equation} 
Then equation (\ref{w1}) is equivalent to the Hamilton system 
\begin{equation}\label{w1ham} 
\dot\psi(t)=D_\pi\cH(\psi(t),\pi(t)), \qquad \dot\pi(t)=-D_\psi\cH(\psi(t),\pi(t)), 
\end{equation} 
(where $\psi(t):=\psi(\cdot,t)$ and $\pi(t):=\pi(\cdot,t)$) with the conserved Hamilton functional 
\begin{equation}\label{ham} 
\cH(\psi,\pi)=\fr12\int[|\pi(x)|^2+|\psi'(x)|^2]\,dx+U(\psi(0)), \qquad (\psi,\pi)\in\cE. 
\end{equation} 
This functional is defined and is G\^ateaux-differentiable on the Hilbert phase space $\cE$. 
We will assume that 
\begin{equation}\label{conf} 
U(\psi)\xrightarrow
[|\psi|\to\infty]{}\infty. 
\end{equation} 
In this case it is easy to prove that the finite energy solution 
$Y(t)=(\psi(t),\pi(t))\in C(\mathbb R,\cE)$ 
exists and is unique for any initial state $Y(0)\in\cE$. Moreover, the solution is bounded:
\begin{equation}\label{solb}
\sup_{x,t\in   \mathbb R}|\psi(x,t)|<\infty.
\end{equation}
We denote $Z:=\{z\in\mathbb R^N:$ $F(z) = 0\} $. Obviously, every 
stationary solution of equation (\ref{w1}) is a~constant function $\psi_z(x)=z\in\mathbb R^N$, 
where $z\in Z$. Therefore, the manifold $\cS$ of all stationary states is a~subset of $\cE$, 
\begin{equation}\label{cS0} 
\cS:=\{S_z=(\psi_z,0): z\in Z\}. 
\end{equation} 
If the set $Z$ is discrete in $\mathbb R^N$, then $\cS$ is also discrete in $\cE$. 
For example, in the case $N=1$
we can consider
the Ginzburg--\allowbreak Lan\-dau potential
$U=(\psi^2-1)^2/4$, 
and respectively, $F(\psi)=-\psi^3 + \psi$. Here the set $Z=\{0, \pm 1\} $ is discrete, and we 
have three stationary states $\psi(x) \equiv 0, \pm 1$. 
\medskip 
 
For $R> 0$ we introduce the following seminorm on the Hilbert phase space 
\begin{equation}\label{cER} 
\Vert(\psi,\pi)\Vert_{\cE_R}=\Vert\psi'\Vert_R+|\psi(0)|+\Vert\pi\Vert_R,\qquad (\psi,\pi)\in\cE, 
\end{equation} 
where $\Vert\cdot \Vert_R$ stands for the norm in $L^2_R:=L^2([-R,R])$. 
We also introduce the following metric on the space $\cE$: 
\begin{equation}\label{metr} 
{\rm dist} [Y_1, Y_2]= 
\sum_1^\infty 2^{-R}\ds\fr{\Vert Y_1-Y_2\Vert_{\cE_R}} 
{1+\Vert Y_1-Y_2\Vert_{\cE_R}}, \quad Y_1,Y_2\in\cE. 
\end{equation} 
The main result of \cite{K1991, K1995a} is the following theorem, which is illustrated with Fig.~\ref{fig-1}. 
 
\begin{theorem}\label{t1} 
{\rm i)} Assume that conditions \eqref{FU} and \eqref{conf}  hold. Then 
\begin{equation}\label{Z} 
Y(t)\xrightarrow
[t\to\pm\infty]{} \cS, 
\end{equation} 
in the metric \eqref{metr} for any finite energy solution $Y(t)=(\psi(t),\pi(t))$. This means that 
\begin{equation}\label{conv} 
{\rm dist} [Y(t), \cS]:= 
\inf_{S\in\cS} {\rm dist} [Y(t), S] 
 \xrightarrow
[t\to\pm\infty]{} 0. 
\end{equation} 
{\rm ii)} Assume, in addition, that $Z$ is a~discrete subset of $\mathbb R^N$. Then 
\begin{equation}\label{Zd} 
Y(t)\xrightarrow
[t\to\pm\infty]{} S_\pm\in \cS, 
\end{equation} 
where the convergence holds in the metric \eqref{metr}.

\begin{figure}[htbp] 
\vspace{-30mm} 
\begin{center} 
\includegraphics[width=0.7\columnwidth]{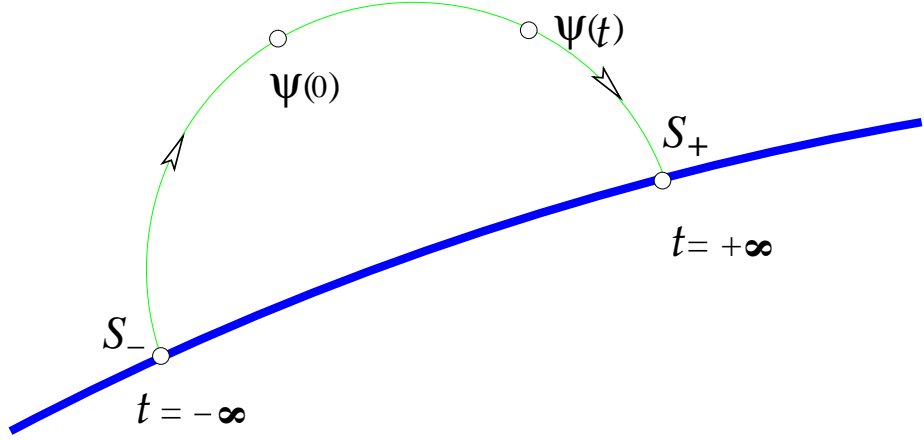} 
\caption{Convergence to stationary states} 
\label{fig-1} 
\end{center} 
\end{figure} 
 
\end{theorem} 
{\bf Sketch of the proof.} 
It suffices to consider only the case $t\to\infty$.
The solution admits the d'Alembert representations for $x> 0$ and $x <0$, which 
imply the `reduced equation' for $y(t):=\psi(0,t)$: 
\begin{equation}\label{rede} 
2\dot y(t)=F(y(t))+2\dot w_{\rm in}(t), \qquad t>0. 
\end{equation} 
Here $w_{\rm in}(t)$ is the sum of incoming waves, for which 
 $\ds\int_0^\infty |\dot w_{\rm in}(t)|^2dt<\infty$. 
This equation provides the `integral of dissipation' 
\begin{equation}\label{reded} 
2\int_0^t |\dot y(s)|^2ds+U(y(t))=U(y(0))+ 2\int_0^t \dot w_{\rm in}(t)\cdot\dot y(s)\,ds, \qquad t>0, 
\end{equation} 
which implies that $\ds\int_0^\infty |\dot y(t)|^2dt<\infty$ according to (\ref{conf}). Hence, (\ref{solb}) implies that
\begin{equation}\label{redy} 
y(t)\to Z, \qquad \dot y(t)\to 0, \qquad t\to\infty. 
\end{equation} 
This convergence 
implies (\ref{Z}), since $\psi(x,t)\sim y(t-|x|)$ for large $t$ and bounded $|x|$. 
\bob
\medskip 
 
Note that the attractions (\ref{Z}) and (\ref{Zd}) in the global norm of $\cE$ is impossible due to outgoing d'Alembert's waves $y(t-|x|)$, 
representing a~solution for large $t$, which carry energy to infinity. In particular, the energy of the limiting stationary state may be smaller that 
the conserved energy of the solution, since the energy of the outgoing waves is irretrievably lost at infinity. 
 Indeed, the energy is the Hamilton functional (\ref{ham}), where the integral vanishes 
 for the limit state, and only the energy of the oscillator $U(\psi(0))$ persists. 
 Therefore, the energy of the limit is usually smaller than the energy of the solution. 
This limit jump is similar to the well-known property of the weak convergence in the Hilbert space. 
 
The discreteness of the set $Z$ is essential: asymptotics (\ref{Zd}) 
can break down if $F(z)=0$ on $[z_-, z_+] $, where $z_- <z_+$. For example,  (\ref{Zd}) breaks down 
for the solution $\psi(x,t)=\sin [\log(|x-t|+2)]$
in the case $z_\pm=\pm 1$. 
\medskip 
 
Further, asymptotics (\ref{Zd}) in the local seminorms can be extended 
to the asymptotics in the global norms (\ref{cE}), taking into account the outgoing d'Alembert's waves. 
Namely, in \cite{KM2009a} we have proved the following result. 
Let us denote by $\cE_*$ the space of $(\phi_0, \pi_0) \in \cE$ for which there exist the finite limits and the integral
(\ref{lim}), and  by $\cE_*^\pm$ the subspace of $\cE_*$ defined by the identity
\begin{equation}\label{idef} 
C_+ + C_- \pm\int_{-\infty}^\infty \pi_0(y)dy=0
\end{equation} 
 in the notations (\ref{lim}).
\begin{theorem}\label{t2} 
Let conditions of Theorem \ref{t1} i) and ii)  hold.
Then for any initial state $(\phi_0, \pi_0) \in \cE_*$
\begin{equation}\label{at4} 
(\psi(\cdot,t),\dot\psi(\cdot,t))=S_{\pm}+W(t)\Phi_\pm+r_{\pm}(t), 
\end{equation} 
where $S_{\pm}\in \cS$, by 
$W(t)$ we denote the dynamical group of the free wave equation \eqref{dal}, $\Phi_\pm \in \cE_*^\pm$ are some `scattering states' of finite energy, and 
the remainder 
$r_{\pm}(t)$ converges to zero in the global energy norm: 
\begin{equation}\label{glob2} 
\Vert r_{\pm}(t)\Vert_{\cE}\xrightarrow
[t\to\pm\infty]{} 0. 
\end{equation} 
\end{theorem} 
The term $W(t)\Phi_\pm$ represents the outgoing d'Alembert's waves, and the condition
(\ref{idef}) provides that the $W(t)\Phi_\pm\to 0$ as $t\to\pm\infty$, according to 
(\ref{dal2}) and (\ref{dal3}).
Thus, Theorem \ref{t2} proves the "Grand Conjecture" \cite[p.460]{soffer2006}
for equation (\ref{w1}).
\medskip 
 
Finally, the asymptotic completeness of this nonlinear scattering was established in \cite{KM2009b, KM2013}.
Let  us fix a stationary state
$S_+= (z_+,0)\in\cS$, and  denote by $\cE_*(S_+)$ the set of initial states 
$(\psi_0, \pi_0)\in \cE_*$ providing the asymptotics (\ref{at4}) with limit state $S_+$ as $t\to\infty$.
Let $F'(z_+)$ denote the corresponding Jacobian matrix and  $\si(F'(z_+))$ denote its spectrum. 

\begin{theorem}\label{t3} 
Let conditions of Theorem \ref{t2} hold.
Then the
 mapping $ (\psi_0, \pi_0) \mapsto \Phi_+$ is the epimorphism $\cE_*(S_+) \to \cE_*^+$ 
if $\rRe \lam \ne 0$ for $\lam\in\si(F'(z_+))$. 
\end{theorem} 

Similar theorem holds obviously for the map $ (\psi_0, \pi_0) \mapsto \Phi_-$.

\subsection{Generalizations} 
{\bf I.} In \cite{K1991, K1995a, KM2009a}, Theorems \ref{t1} and \ref{t2} were established also for more general equation than  
(\ref{w1}):
\begin{equation}\label{w1m} 
(1+m\de(x))\ddot\psi(x,t)=\psi''(x,t) + \delta(x) F(\psi(0, t)), \qquad x \in \mathbb R, 
\end{equation} 
where $m>0$ is the mass of the particle attached to the string at the point $x=0$.
In this case the Hamiltonian (\ref{ham}) includes the additional term $mv^2/2$, where $v=\dot\psi(0,t)$.
Moreover, the reduced equation (\ref{rede}) now becomes
the Newton equation with the friction:
\begin{equation}\label{redem} 
m\ddot y(t)=
F(y(t))-2\dot y(t)+2\dot w_{\rm in}(t), \qquad t>0. 
\end{equation} 
{\bf II.} In \cite{K1995b}, we have proved the convergence (\ref{Z}) and (\ref{Zd}) to a~global attractor 
for the string with $N$ oscillators: 
\begin{equation}\label{wN} 
\ddot\psi(x,t)=\psi''(x,t)+\sum_1^N\delta(x-x_k)F_k(\psi(x_k,t)). 
\end{equation} 
The equation is reduced to a~system of $N$ equations with delay, 
but its study requires novel arguments, since the oscillators are connected at different moments of time. 
 \medskip\\ 
{\bf III.} In \cite{K1999}, the result was extended to equations of the type 
\begin{equation}\label{neli} 
\ddot\psi(x,t)=\psi''(x,t)+\chi(x)F(\psi(x,t)), 
\end{equation} 
where $\chi\in C_0^\infty(\mathbb R)$,  $\chi(x)\ge 0$, and  $\chi(x)\not\equiv 0$  while 
$F$  has structure (\ref{FU}) with potential $U$ satisfying (\ref{conf}). 
This guarantees the existence of global solutions of finite energy and conservation of the Hamilton functional 
\begin{equation}\label{hama} 
\cH(\psi,\pi)=\fr12\int[|\pi(x)|^2+|\psi'(x)|^2+\chi(x)U(\psi(x))]\,dx. 
\end{equation} 
{\bf Sketch of the proof.} Again
it suffices to consider only the case $t\to\infty$.
For the proof of (\ref{Z}) and (\ref{Zd}) in this case  we 
develop our approach \cite{K1995b} based on the finiteness of energy radiated from an interval $[-a, a]\supset\supp\chi$, which 
implies the finiteness of `integral of dissipation' \cite[(6.3)]{K1999}: 
\begin{equation}\label{idis} 
\int[|\dot\psi(-a,t)|^2+|\psi'(-a,t)|^2+|\dot\psi(a,t)|^2+|\psi'(a,t)|^2]dt<\infty. 
\end{equation} 
This means, roughly speaking, that 
\begin{equation}\label{idis2} 
\psi(\pm a,t)\sim C_\pm,\qquad \psi'(\pm a,t)\sim 0, \qquad t\to\infty. 
\end{equation} 
It remains to justify the correctness of the boundary value problem for nonlinear differential equation (\ref{neli}) 
in the band $-a \le x \le a$, $t> 0$, with the Cauchy boundary conditions (\ref{idis2}) on the sides $x = \pm a$. 
This correctness should imply the convergence of type 
\begin{equation}\label{idis3} 
\psi(x,t)\sim S(x), \qquad t\to\infty. 
\end{equation} 
The proof employs the symmetry of the wave equation with respect to permutations of variables 
 $x$ and $t$ with simultaneous change of sign of the potential $U$. In this boundary-value problem 
 the variable $x$ plays the role of time, and 
condition (\ref{conf}) makes the potential unbounded from below! Hence, 
this dynamics with $x$ as `time variable' is not globally correct on the interval $|x| \le a$: 
for example, in the ordinary equation 
 $\psi''(x)-U'(\psi)=0$ with $U=\psi^4$, 
 a~solution can run away at a point  $x\in(-a,a)$. However, in our setting the local correctness is sufficient 
in view of the \textit{a~priori} estimates, 
which follow from the conservation of energy (\ref{hama}) 
due to the conditions (\ref{conf}) and $\chi(x)\ge 0$, $\chi(x)\not\equiv 0$. \bob
\medskip\\ 
A detailed presentation of the results \cite{K1991}--\cite{K1999} is available in the survey \cite{K2000}.

 \subsection{Wave-particle system} 
In \cite{KSK1997} we have proved the first result on the global attraction 
\eqref{ate} for the 3-dimensional 
real scalar wave field coupled to a~relativistic particle. 
The 3D scalar field satisfies the wave equation 
\begin{equation}\label{w3} 
\ddot\psi(x,t)=\Delta\psi(x,t)-\rho(x-q(t)), \qquad x\in\mathbb R^3, 
\end{equation} 
where $\rho\in C^\infty_0(\mathbb R^3)$ is a~fixed function, representing the charge density of the particle, 
and $q(t)\in\mathbb R^3$ is the particle position. The particle motion obeys the Hamilton equations with the 
relativistic kinetic energy $\sqrt{1+p^2}$: 
\begin{equation}\label{q3} 
\dot q(t)=\frac{p(t)}{\sqrt{1+p^2(t)}}, \qquad \dot p(t)=-\nabla V(q(t))-\int\nabla\psi(x,t)\rho(x-q(t))\,dx. 
\end{equation} 
Here, $-\nabla V(q)$ is the external force produced by some real potential $V(q) $, and the integral is the self-force. This means that the 
wave function $\psi$, generated by the particle, 
plays the role of a~potential acting on the particle, along with the external potential $V(q)$. 
 
\begin{definition} 
{\rm $\cE:=H^1(\mathbb R^3)\oplus L^2(\mathbb R^3)\oplus\mathbb R^3\oplus\mathbb R^3$ is the Hilbert phase space of 	 
tetrads $(\psi,\pi,q,p)$ with finite norm 
\begin{equation}\label{cE3} 
\Vert(\psi,\pi,q,p)\Vert_{\cE}=\Vert\na\psi\Vert+\Vert\psi\Vert+\Vert\pi\Vert+|q|+|p|, 
\end{equation} 
where $\Vert\cdot\Vert$ is the norm in $L^2:= L^2(\mathbb R^3)$}. 
 \end{definition} 
 
System \eqref{w3}--\eqref{q3} is equivalent to the Hamilton system 
\begin{equation}\label{w3ham} 
\left\{ 
\begin{array}{rclrcl} 
\dot\psi(t)&=&D_\pi\cH(\psi(t),\pi(t),q(t),p(t)), &\qquad  \dot\pi(t)&=&-D_\psi\cH(\psi(t),\pi(t),q(t),p(t))\\ 
\\ 
\dot q(t)&=&D_p\cH(\psi(t),\pi(t),q(t),p(t)),  & \qquad \dot p(t)&=&-D_q\cH(\psi(t),\pi(t),q(t),p(t)) 
\end{array}\right| 
\end{equation} 
with the conserved Hamilton functional 
\begin{equation}\label{Ham} 
\cH(\psi,\pi, q,p)\!=\!\ds\frac12\!\int[|\pi(x)|^2\!+\!|\na\psi(x)|^2]\,dx\!+\!\int\psi(x)\rho(x\!-\!q)\,dx\!+\! 
\sqrt{1\!+\!p^2}\!+\!V(q), \quad (\psi,\pi,q,p)\in\cE. 
\end{equation} 
This functional is defined and is G\^ateaux-differentiable 
on the Hilbert phase space $\cE$. 
\medskip 
 
We assume that the potential 
 $V(q)\in C^2(\mathbb R^3)$ is confining: 
\begin{equation}\label{V} 
V(q)\xrightarrow
[|q|\to\infty]{}\infty. 
\end{equation} 
In this case 
it is easy to prove that the finite energy solution 
$Y(t)=(\psi(t),\pi(t),q(t),p(t))\in C(\mathbb R,\cE)$ 
exists and is unique for any initial state  $Y(0)\in\cE$. 
 
In the case of a~point particle $\rho(x)=\delta(x)$ the system \eqref{w3}--\eqref{q3} 
is undetermined. Indeed, in this setting any solution to the wave equation \eqref{w3} is singular at $x=q(t)$, 
and respectively, the integral on the right of \eqref{q3} does not exist. 
 
We denote $Z =\{z\in\mathbb R^3: $ $ \na V(z) = 0\}$. 
It is easily checked that the stationary states of the system \eqref{w3}--\eqref{q3} are of the form 
\begin{equation}\label{statsol} 
S_z=(\psi_z,0,z,0), 
\end{equation} 
where  $z\in Z$, while $\De\psi_z (x)=\rho(x-z)$; i.e., 
$$ 
\psi_z (x):=-\ds\fr1{4\pi}\int \fr{\rho(y-z)\,dy}{|x-y|} 
$$ 
is the Coulomb potential. 
Respectively, 
the set of all stationary states of this system is given by 
\begin{equation}\label{cS03} 
\cS:=\{S_z: z\in Z \}. 
\end{equation} 
 
If the set $Z\subset \mathbb R^N$ is discrete, then $\cS $ is also discrete in $\cE$. 
Finally, we 
assume that the `form factor' $\rho$ satisfies the Wiener condition 
\begin{equation}\label{W1} 
\hat\rho(k):=\int e^{ikx}\rho(x)\,dx\ne 0, \qquad k\in\mathbb R^3. 
\end{equation} 
It means the strong coupling of the scalar field $\psi(x)$ with the particle.

Let us denote $B_R=\{x\in\mathbb R^3: |x|<R\}$ for $R>0$ and let $\Vert\cdot\Vert_R$ stand for the norm in $L^2(B_R)$. 
We define the local energy seminorms 
\begin{equation}\label{cER3} 
\Vert(\psi,\pi,q,p)\Vert_{\cE_R}=\Vert\na\psi\Vert_R+\Vert\psi\Vert_R+\Vert\pi\Vert+|q|+|p| 
\end{equation} 
on the Hilbert phase space $\cE$. 
The main result of \cite{KSK1997} is the following. 
 
\begin{theorem}\label{t4} 
{\rm i)} Let conditions \eqref{V}, \eqref{W1} hold, and let $Y(t)=(\psi(t),\pi(t),q(t),p(t))$ 
be a~finite energy solution to the system \eqref{w3}--\eqref{q3}. Then 
\begin{equation}\label{Z3} 
Y(t)\xrightarrow
[t\to\pm\infty]{}\cS, 
\end{equation} 
where the convergence holds in the metric \eqref{metr} with seminorm \eqref{cER3}. 
\medskip\\ 
{\rm ii)} Let moreover, the set  $Z$ be discrete in~$\mathbb R^N$. Then 
\begin{equation}\label{Zd3} 
Y(t)\xrightarrow
[t\to\pm\infty]{}S_\pm\in \cS, 
\end{equation} 
where the convergence holds in the same metric. 
 \end{theorem} 
 \textbf{Sketch of the proof.} 
 The key point in the proof 
 is the relaxation of acceleration 
\begin{equation}\label{rel} 
\ddot q(t)\xrightarrow
[t\to\pm\infty]{} 0,
\end{equation} 
which follows from the Wiener condition \eqref{W1}. Then
the asymptotics \eqref{Z3} and \eqref{Zd3} immediately follow from
this relaxation and from \eqref{V} by the Li\'enard-Wiechert representations for the potentials.

Let us explain  
how to deduce (\ref{rel}) 
as  $t\to\infty$
in the case 
of spherically symmetric 
form factor $\rho(x)=\rho_1(|x|)$.
The energy conservation and condition \eqref{V} imply the \textit{a~priori} estimate $|p(t)| \le \co $, and hence 
\begin{equation}\label{dqapr} 
|\dot q(t)|\le \ov v<1 
\end{equation} 
by the first equation of \eqref{q3}. The radiated energy during the time $0 <t <\infty$ 
is finite by condition \eqref{V}:
\begin{equation}\label{dissinf} 
E_{\rm rad}=\lim_{R\to\infty}\int_0^\infty \Big[\int_{|x|=R} S(x,t)\cdot \fr x{|x|}d^2x\Big] dt<\infty, 
\end{equation} 
where $S(x,t)=-\pi(x,t) \na \psi (x,t)$ is the density of energy flux.
Let us denote 
\begin{equation}\label{dissipR} 
R_\om(t):=
\int \rho(y-q(t+\om\cdot y)) \fr{\om\cdot \ddot q(t+\om\cdot y)}
{[1-\om\cdot \dot q(t+\om\cdot y)]^2}dy, \qquad \om\in\mathbb R^3, \, |\om|=1. 
\end{equation} 
It turns out that the finiteness of energy radiation (\ref{dissinf}) 
also implies the finiteness of the integral 
\begin{equation}\label{dissip} 
I_{\rm rad}=\int_0^\infty  \Big[\int_{|\om|=1}|R_\om(t)|^2 \5 d^2\om\Big]dt <\infty, 
\end{equation} 
which represents the contribution of the Li\'enard--Wiechert retarded potentials. 
 Furthermore, the function 
$R(\om,t)$ is globally Lipschitz in view of (\ref{dqapr}). Hence, 
\begin{equation}\label{dqaprR} 
R_\om(t)\xrightarrow
[t\to\infty]{} 0,\quad |\om|=1. 
\end{equation} 
To deduce (\ref{rel}), it is necessary to rewrite (\ref{dissipR}) as a~convolution. 
We denote $r(s):=\om\cdot q(s)$ and observe that the map $s\mapsto \theta:= s-r(s)$ 
is a~diffeomorphism from $\mathbb R$ to~$\mathbb R$, inasmuch as $|\dot r(s)| \le \ov v< 1 $ by (\ref{dqapr}). 
Then the desired convolution representation reads 
\begin{equation}\label{Rsv} 
R_\om(t)=[\rho_a*g_\om](t):= \int\rho_a(t-\theta)g_\om(\theta)\5 d\theta,\qquad \rho_a(q_1):=\int dq_2dq_3\rho(q_1,q_2,q_3), 
\end{equation} 
where 
\begin{equation}\label{gom} 
g_\om(\theta):= [1-\dot r(s(\theta))]^{-3}\ddot r(s(\theta)), \qquad \theta\in\mathbb R. 
\end{equation} 
It remains to note that 
$[\rho_a*g_\om](t)\to 0$ by (\ref{dqaprR}), while the Fourier transform $\ti\rho_a (k)\ne 0$ for 
$k\in\mathbb R$ by~\eqref{W1}. 
Now (\ref{rel}) follows from the Wiener Tauberian theorem. \bob
\medskip 
 
In \cite{KSK1997} we have also proved the asymptotic stability of stationary states $S_z$ 
with positive Hessian $d^2V(z)\nobreak >0$.

  \begin{remark}\label{WC1}
{\rm 
i) The proof of relaxation~\eqref{rel} does not depend on the  condition \eqref{V}.
In particular, ~\eqref{rel} holds for $V= 0$.
\medskip\\
ii) 
The Wiener condition~\eqref{W1} is sufficient for the relaxation~\eqref{rel}
for solutions to the system ~\eqref{w3}--\eqref{q3}.
However, it is not  
necessary
for some specific classes of potentials and solutions
in the case of 
 small $\Vert\rho\Vert$, see Section 4.3. }

 \end{remark}

\subsection{Maxwell-Lorentz equations: radiation damping} 
 
In \cite{KS2000} the attractions (\ref{Z3}), (\ref{Zd3}) were extended to the 
Maxwell equations in $\mathbb R^3$ coupled to a relativistic particle: 
\begin{equation}\label{ML} 
\!\! 
\left\{\!\! 
\begin{array}{l} 
\dot E(x,t)\!=\!\rot B(x,t)-\dot q\rho(x\!-\!q),\  \dot B(x,t)\!=\!- \rot E(x,t),  \ \div E(x,t)\!=\!\rho(x\!-\!q),  \ \div B(x,t)\!=\!0 
\\ 
\\ 
\dot q(t)\!=\!\ds\frac{p(t)}{\sqrt{1\!+\!p^2}(t)},\qquad\,\, 
\dot p(t)\!=\!\ds\int[E(x,t)\!+\!E^{\rm ext}(x)\! +\!\dot q(t)\times (B(x,t)\!+\!B^{\rm ext}(x))]\rho(x\!-\!q(t))\,dx 
\end{array}\!\!\right| 
\end{equation} 
where 
$\rho(x-q)$ is the charge density of a~particle,  $\dot q\rho(x-q)$ is the corresponding current density, and 
$E^{\rm ext}(x)= -\na \phi^{\rm ext}(x)$, $B^{\rm ext}(x)= -\rot A^{\rm ext}(x)$ 
are external Maxwell fields. Similarly to 
\eqref{V}, we assume that 
 \begin{equation}\label{VML} 
V(q):=\int \phi^{\rm ext}(x)\rho(x-q)\,dx \xrightarrow
[|q|\to\infty]{}\infty. 
\end{equation} 
This system describes the classical electrodynamics of an 
`extended electron' introduced by Abraham \cite{A1902, A1905}. 
In the case of a~point electron, when $\rho(x)=\delta(x)$, such 
a~system is undetermined. Indeed, in this setting any solutions 
$E(x,t)$ and $B(x,t)$ to the Maxwell equations (the first line of \eqref{ML}) are singular at $x=q(t)$, 
and respectively, the integral on the right of the last equation in \eqref{ML} does not exist. 
 
The system \eqref{ML} is time reversible in the following sense: if 
$E(x,t)$, $B(x,t)$, $q(t)$, $p(t)$ is its solution, then 
$E(x,-t)$, $-B(x,-t)$, $q(-t)$, $-p(-t)$ is also the solution to \eqref{ML} with external fields 
$E^{\rm ext}(x)$, $-B^{\rm ext}(x)$. This system can be 
represented in the Hamilton form if the fields are expressed via the potentials 
 $E(x,t)= -\na \phi(x,t)-\dot A(x,t)$, $B(x,t)= -\rot A(x,t)$. 
 The corresponding Hamilton functional is as follows 
\begin{equation}\label{HAM} 
\cH=\frac12[\langle E, E\rangle+\langle B, B\rangle ]+V(q)+\sqrt{1+p^2}= 
\frac12\int[E^2(x)+B^2(x)]\,dx+V(q)+\sqrt{1+p^2}. 
\end{equation} 
This Hamiltonian is conserved, since 
\begin{eqnarray}\label{HAMc} 
\dot\cH(t)&=&\langle E(x,t),\dot E(x,t) \rangle +\langle B(x,t),\dot B(x,t) \rangle+\na V(q)\cdot\dot q(t)+\dot q(t)\cdot\dot p(t) 
\nonumber\\ 
\nonumber\\ 
&=&\langle E(x,t),\rot B(x,t)-\dot q(t)\rho(x-q(t))\rangle -\langle B(x,t),\rot E(x,t)\rangle-\langle E^{\rm ext}(x),\rho(x-q(t))\rangle\cdot \dot q(t) 
\nonumber\\ 
\nonumber\\ 
&& 
+ 
\dot q(t)\cdot 
\langle E(x,t)+E^{\rm ext}(x) +\dot q(t)\times (B(x,t)+B^{\rm ext}(x)),\rho(x-q(t))\rangle 
\nonumber\\ 
\nonumber\\ 
&=&\langle E(x,t),\rot B(x,t)\rangle-\langle B(x,t),\rot E(x,t)\rangle=-\lim_{R\to\infty}\int_{|x|<R} \div [ E(x,t)\times B(x,t)] dx 
\nonumber\\\ 
\nonumber\\\ 
&=&-\lim_{R\to\infty}\int_{|x|=R} [E(x,t)\times B(x,t)]\cdot\fr x{|x|}\,dS(x)=0. 
\end{eqnarray} 
This energy conservation gives \textit{a~priori} estimates of solutions, which play an important role in the proof of the attractions of type 
(\ref{Z3}), (\ref{Zd3}) 
in \cite{KS2000}. 
The key role in these proofs  
again plays the relaxation of the acceleration
(\ref{rel}) which follows by a suitable development of our methods \cite{KSK1997}:
an expression of type (\ref{dissip}) 
for the radiated energy via the Li\'enard-Wiechert retarded potentials,
the convolution representation of type  (\ref{Rsv}), and the application of the 
Wiener Tauberian theorem.
\medskip

In Classical Electrodynamics
the relaxation \eqref{rel}
is known as the {\bf radiation damping}.
It is  traditionally justified by the  Larmor and Li\'enard  formulas
\cite[(14.22)]{Jackson} and \cite[(14.24)]{Jackson}
for the power of radiation of a point particle. 
These formulas are deduced  from the 
Li\'enard-Wiechert expressions for the retarded potentials
neglecting the initial field and the  "velocity field".
Moreover, the traditional approach neglects the back field-reaction 
though it should be the key reason for the relaxation.
The main problem is that  this back field-reaction  is infinite for the  point particles.
The rigorous meaning to these calculations
has been  suggested first in {\rm \cite{KSK1997, KS2000}} 
for the Abraham model of the `extended electron'
under the Wiener condition \eqref{W1}. 
The survey can be found in   \cite{S2004}.
 
\begin{remark} 
{\rm
All the above results on the attraction of type \eqref{ate} 
relate to `generic' systems with the trivial symmetry group, which are characterized 
by the discreteness of attractors, the Wiener condition, etc. }
\end{remark} 
 
 \medskip

\setcounter{equation}{0} 
\section{Global attraction to stationary orbits} 
The global attraction to stationary orbits (\ref{atU}) was first proved in 
\cite{K2003, KK2006,KK2007} for the Klein--\allowbreak Gordon equation 
coupled to the nonlinear oscillator 
\begin{equation} \label{KG1} 
\ddot \psi(x,t)=\psi''(x,t)-m^2\psi(x,t)+\delta(x) F(\psi(0,t)),\qquad x\in\mathbb R. 
\end{equation} 
We consider complex solutions, identifying $\psi\in\mC$ with $(\psi_1,\psi_2)\in\mathbb R^2$, 
where $\psi_1=\rRe\psi$, $\psi_2=\rIm\psi$. 
We assume that $F\in C^1(\mathbb R^2, \mathbb R^2)$ and 
\begin{equation}\label{C1} 
F(\psi)=-\na_{\ov\psi} U(\psi), \qquad \psi\in\mC, 
\end{equation} 
where $U$ is a~real function, and  $\na_{\ov\psi}:=\pa_1+i\pa_2$. 
In this case equation \eqref{KG1} 
is a~Hamilton system of form (\ref{w3ham}) with the Hilbert 
phase space $\cE:=H^1(\mathbb R) \oplus L^2 (\mathbb R) $ and the conserved 
Hamilton functional 
\begin{equation}\label{U} 
\cH(\psi,\pi)=\frac12\int\Big[ 
 |\pi(x)|^2+|\psi'(x)|^2 
+m^2|\psi(x)|^2\Big]\,dx + U(\psi(0)),\qquad (\psi,\pi)\in\cE.
\end{equation} 
We assume that 
 \begin{equation}\label{C2} 
 \inf_{\psi\in\mC} U(\psi)>-\infty. 
 \end{equation} 
 In this case a~finite energy solution  $Y(t)=(\psi(t),\pi(t))\in C(\mathbb R,\cE)$ 
exists and is unique for any initial state $Y(0)\in\cE$. 
The  \textit{a~priori} estimate  
\begin{equation}\label{apri} 
 \sup_{t\in\mathbb R} [\Vert \pi(t)\Vert_{L^2(\mathbb R)}+\Vert \psi(t)\Vert_{H^1(\mathbb R)}]<\infty 
\end{equation} 
holds due to the conservation of Hamilton functional (\ref{U}). 
Note that condition \eqref{conf} now is not necessary, since the conservation of functional (\ref{U}) with $m> 0$ provides 
 boundedness of the solution. 
 
 Further, we assume the $ U (1) $-invariance of the potential: 
\begin{equation}\label{C3} 
U(\psi)=u(|\psi|), \qquad \psi\in\mC. 
\end{equation} 
Then the differentiation (\ref{C1}) gives
\begin{equation}\label{Fap} 
F(\psi)=a(|\psi|)\psi,\qquad \psi\in \mC, 
\end{equation} 
and hence, 
\begin{equation}\label{U1} 
F(e^{i\theta}\psi)=e^{i\theta}F(\psi), \qquad \theta\in\mathbb R. 
\end{equation} 
 
By  `stationary orbits' (or solitons) we shall understand any solutions of the form 
$\psi_\om(x,t)=\phi_\om(x)e^{-i\omega t}$ with 
$\phi_\om\in H^1(\R)$ and 
$\om\in\R$. Each stationary orbit provides the 
corresponding solution to the \textit{nonlinear eigenfunction problem} 
\begin{equation}\label{sw} 
 -\om^2 \phi_\om(x)=\phi''_\om(x)-m^2\phi_\om(x)+\delta(x) 
F(\phi_\om(0)), \qquad x\in\mathbb R. 
\end{equation} 
The solutions 
$\phi_\om\in H^1(\R)$  
have the form 
$\phi_\om(x)=Ce^{-\ka|x|}$, where $\ka:=\sqrt{m^2-\om^2}>0$ and $C$ satisfies the equation 
$$ 
2\ka C=F(C). 
$$ 
Hence, the solutions exist for $\om\in\Om$, where 
$\Om$ is a~subset of the \textit{spectral gap} $[-m,m]$. Let us define the corresponding 
\textit{solitary manifold} 
\begin{equation}\label{Sol} 
{\cal S}=\{(e^{i\theta} \phi_\om,-i\om e^{i\theta} \phi_\om)\in\cE :\om\in\Om, ~\theta\in[0,2\pi]\}. 
\end{equation} 

Finally, we assume that equation 
\eqref{KG1} is \textit{strictly nonlinear:} 
\begin{equation}\label{C4} 
U(\psi)=u(|\psi|^2)=\Si_0^{N}u_{j}|\psi|^{2j},\quad u_N>0, \quad N\ge 2. 
\end{equation} 
For example, the known \textit{Ginzburg--\allowbreak Landau potential} 
$U(\psi)=|\psi|^4/4-|\psi|^2/2$ satisfies all conditions \eqref{C2}, \eqref{C3} and \eqref{C4}. 
 
\begin{figure}[htbp] 
\begin{center} 
\includegraphics[width=0.9\columnwidth]{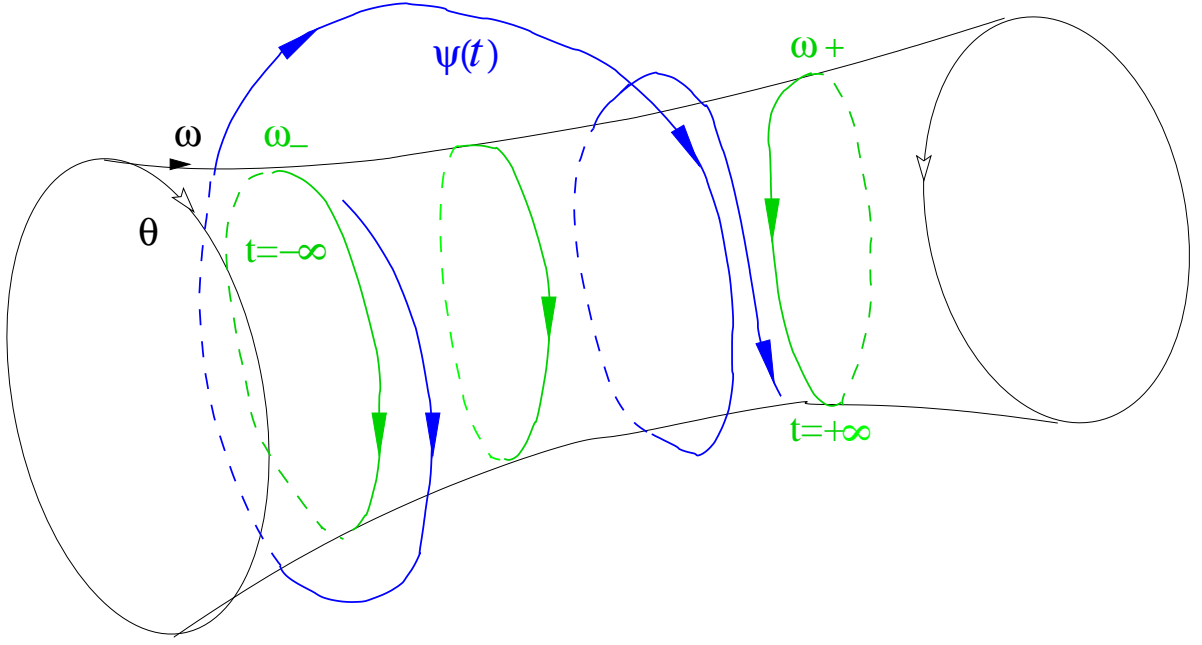} 
\caption{Convergence to stationary orbits} 
\label{fig-3} 
\end{center} 
\end{figure}

\begin{theorem}\label{t5} 
Let conditions \eqref{C1}, \eqref{C2}, \eqref{C3} and \eqref{C4} hold. Then 
any finite energy solution $Y(t)=(\psi(t),\pi(t))$ 
to equation \eqref{KG1} converges to the solitary manifold 
in the long time limits {\rm (}see Fig.~{\rm \ref{fig-3}):} 
\begin{equation}\label{ga} 
Y(t)\xrightarrow
[t\to\pm\infty]{} {\cal S}, 
\end{equation} 
where the convergence holds in the sense of \eqref{conv}. 
\end{theorem} 
 
\textbf{Generalizations:} 
Attraction (\ref{ga}) is extended in \cite{KK2010a} to the 1D Klein--\allowbreak Gordon equation with $N$ nonlinear oscillators 
\begin{equation}\label{KGN} 
\ddot \psi(x,t)=\psi''(x,t)-m^2\psi+ 
\sum\limits_{k=1}^N\delta(x-x_k)F_k(\psi(x_k,t)),~x\in \mathbb R, 
\end{equation} 
and in \cite{KK2009, KK2010b}, to the \textit{n}D Klein--\allowbreak Gordon and Dirac equations with a~`nonlocal interaction' 
\begin{eqnarray} 
\ddot \psi(x,t)&=&\De\psi(x,t)-m^2\psi+ 
\rho(x)F(\langle\psi(\cdot,t),\rho\rangle), \qquad 
x\in \mathbb R^n,\label{KGn}\\ 
\nonumber\\ 
i\dot\psi(x,t) 
&=& 
\big(-i\alpha\cdot\na +\beta m\big) \psi 
+\rho(x)F(\langle\psi(\cdot,t),\rho\rangle), \qquad x\in \mathbb R^n \label{Dn} 
\end{eqnarray} 
under the Wiener condition \eqref{W1}, where $\alpha = (\alpha_1, \dotsc , \alpha_n)$ and $\beta=\al_0$ are the Dirac matrices. 
\medskip\\ 
Furthermore, attraction  (\ref{ga}) is extended in \cite{C2013} to discrete in space and time nonlinear Hamilton equations, which are discrete approximations of equations like (\ref{KGn}). 
The proof relies on the new refined version of the Titchmarsh theorem for distributions on the circle, as obtained in~\cite{KK2013}. 
 
\newpage
\textbf{Open questions:} 
 \medskip\\ 
I. Attraction (\ref{atU}) to the orbits with fixed frequencies $\om_\pm$. 
\medskip\\ 
II. Attraction to stationary orbits (\ref{ga}) for nonlinear Schr\"odinger equations. In particular, 
for the 1D Schr\"odinger equation coupled to a~nonlinear oscillator 
\begin{equation} \label{S1} 
i\dot \psi(x,t)=-\psi''(x,t)+\delta(x) F(\psi(0,t)), \qquad x\in\mathbb R 
\end{equation} 
(see Remark \ref{rST}). 
\medskip\\ 
III. Attraction to solitons \eqref{att} for the \textbf{relativistically-invariant} nonlinear Klein--\allowbreak Gordon equations. 
In particular, for the 1D equations 
$$ 
\ddot\psi(x,\!t)=\psi''(x,t)-m^2\psi(x,t)+F(\psi(x,t)). 
$$ 

 Below we give  a~schematic proof of Theorem \ref{t5} 
 in a more simple case of the zero initial data: 
\begin{equation}\label{ini0} 
\psi(x,0)=0,\qquad \dot\psi(x,0)=0. 
\end{equation} 
The general case of nonzero initial data is reduced 
to (\ref{ini0})
by a trivial subtraction \cite{K2003,KK2007}.
The proof relies on a~new strategy, which was first introduced in \cite{K2003} and refined in~\cite{KK2007}. 
The main steps of the strategy are the following: 
 \medskip\\  
(1) The Fourier transform in time for finite energy solutions to the nonlinear equation (\ref{KG1}). 
 \smallskip\\
(2) Absolute continuity of the Fourier transform 
on the continuous spectrum of the free Klein-Gordon equation. 
 \vspace{-4mm}
 \smallskip\\
(3) The reduction of spectrum of omega-limit trajectories to a~subset of the corresponding spectral gap. 
 \smallskip\\
(4) The reduction of this subset  to a~single point. 
\medskip\\ 
The steps (2) and (4) are central in the proof. 
The property (2) is a~nonlinear analog of the Kato Theorem on the absence of embedded eigenvalues in 
the continuous spectrum; it implies (3). 
Step (4) is justified by the Titchmarsh convolution theorem. It means that the limiting behavior 
of any finite energy solution is single-frequency, which essentially coincides with asymptotics (\ref{atU}). 
An important technical role plays the application of the theory of quasi-measures and their multipliers \cite[Appendix B]{KK2007}. 
\medskip 
 
The strategy (1)--(4) was also employed in \cite{KK2009}--\cite{C2013}. 
 
\subsection{Spectral representation and quasi-measures} 
It suffices to prove attraction (\ref{ga}) only for positive times: 
\begin{equation}\label{gap} 
Y(t)\xrightarrow
[t\to +\infty]{} {\cal S}, 
\end{equation} 
We extend $\psi(x,t)$ and $f(t):=F(\psi (0,t))$ by zero for $t <0$ and denote 
\begin{equation}\label{gap2} 
\psi_+(x,t):= 
\left\{ 
\begin{array}{ll} 
\psi(x,t),&t>0,\\ 
0,& t<0, 
\end{array}\right. 
\qquad f_+(t):= 
\left\{ 
\begin{array}{ll} 
f(t),&t>0,\\ 
0,& t<0. 
\end{array}\right. 
\end{equation} 
By \eqref{KG1} and (\ref{ini0}) these functions satisfy the following equation 
\begin{equation}\label{KG2} 
\ddot \psi_+(x,t)=\psi_+''(x,t)-m^2\psi_+(x,t)+\delta(x) f_+(t), \qquad (x,t)\in\mathbb R^2 
\end{equation} 
in the sense of distributions. 
We denote by $\ti g(\om)$ the Fourier transform of the tempered distribution $g(t)$ given by 
\begin{equation}\label{Fu} 
\ti g(\om)=\int_\mathbb R e^{i\om t} g(t)\,dt,\qquad \om\in\mathbb R 
\end{equation} 
for test functions $g \in C_0^\infty (\mathbb R)$. 
It is important that $\psi_+(x,t) $ and $_+f(t)$ are bounded functions of $t\in\mathbb R$ 
with values 
in the Sobolev space $H^1(\mathbb R)$ and $\mC$, respectively, 
due to the \textit{a~priori} estimate~\eqref{apri}. 
Now the Paley--Wiener theorem \cite[p.~161]{K1999} implies that 
their Fourier transforms admit an extension from the real axis to an analytic 
functions of $\om\in\mC^+:=\{\om\in\mC:~\rIm\om>0\}$ with values in $H^1(\mathbb R)$ and $\mC $, respectively: 
\begin{equation}\label{FL} 
\ti\psi_+(x,\om)=\int_0^\infty e^{i\om t}\psi(x,t)\,dt,\qquad \ti f_+(\om)=\int_0^\infty e^{i\om t}f(t)\,dt,\qquad \om\in\mC^+. 
\end{equation} 
These functions grow not faster than $|\rIm \om|^{-1}$ as 
$\rIm \om \to 0+$ 
in view of (\ref{apri}). Hence, 
their boundary values at $\om \in \mathbb R$ are the distributions of a~low singularity:
they are second-order derivatives of continuous functions as in the case $\ti f_+(\om)=i/(\om-\om_0)$ with $\om_0\in\R$,
which corresponds to 
$f_+(t)=\theta(t)e^{-i\om_0 t}$. 

Recall that the Fourier transform of functions from $L^\infty(\mathbb R)$ are called quasi-measures 
 \cite{Gaudry1966}. 
 Further we will use a~special weak `Ascoli--\allowbreak Arzela' convergence in the space $L^\infty(\mathbb R)$: 
 
\begin{definition}\label{dAA} 
{\rm For $g, g_n \in L^\infty (\mathbb R)$ the convergence $g_n \toAA g$ means that 
\begin{equation}\label{AA} 
\lim_{n\to\infty} 
\Vert g_n(t)-g(t)\Vert_{L^\infty(-T,T)}= 0\ \ \ \forall T>0\quad \mbox{and}\quad \sup_{n} \Vert g_n\Vert_{L^\infty(\mathbb R)}<\infty. 
\end{equation} 
} 
\end{definition} 
 
\begin{definition}\label{dQMA} 
{\rm i) A~tempered distribution $\mu (\om)$ is called a~\textit{quasi-measure} if $\mu=\ti g$, where $g\in L^\infty (\mathbb R) $. 
 
ii) $\cQ \cM$ denotes the linear space of quasi-measures endowed with the following convergence: 
for a~sequence 
 $\mu_n=\ti g_n\in \cQ\cM$ with $g_n\in L^\infty(\mathbb R)$ 
\begin{equation}\label{QMA} 
\mu_n \toQM \mu \quad \mbox{if and only if}\quad g_n \toAA g. 
\end{equation} 
} 
\end{definition} 
The following technical lemma will play 
an important role in our analysis. Denote $L^1:= L^1(\mathbb R)$. 
 
\begin{lemma}\label{lQM} 
{\rm i)} The function $M(\om)$ is a~multiplier in $\cQ \cM$ if 
 $M=\ti G$, where $G\in L^1$. 
 
{\rm ii)} Let $\mu_n \toQM \mu$, and $ G_n\toL G$. Then, for $M_n:=\ti G_n$ and $M=\ti G$, 
\begin{equation}\label{QMA2} 
M_n\mu_n \toQM M \mu. 
\end{equation} 
\end{lemma} 
For the proof it suffices to verify that $G_n*g_n \toAA G*g $ if $g_n \toAA g$. 
\medskip\\ 
Further, by \eqref{ini0} equation 
(\ref{KG2}) in the Fourier transform reads as the stationary Helmholtz equation 
\begin{equation}\label{KG3} 
-\om^2 \ti\psi_+(x,\om)=\ti\psi_+''(x,\om)-m^2\ti\psi_+(x,\om)+\delta(x) \ti f_+(\om),\qquad x\in\mathbb R. 
\end{equation} 
Its solution is given by 
\begin{equation}\label{KG4} 
\ti\psi_+(x,\om)=-\ti f_+(\om)\fr{e^{ik(\om)|x|}}{2ik(\om)},\qquad \rIm\om>0. 
\end{equation} 
Here $k(\om):=\sqrt{\om^2-m^2}$, 
where the 
branch of the root is chosen to be analytic for $\rIm \om> 0$ and having positive imaginary part. 
For this branch, the right-hand side of equation (\ref{KG4}) belongs to $H^1(\mathbb R) $ 
in accordance with the properties of $\ti \psi_ + (x,\om)$, while 
for the other branch the right-hand side grows exponentially as $|x| \to \infty$. 
Such argument for the choice of the solution is known as the `limiting absorption principle' 
in the theory of diffraction \cite{KopK2012-1}. 
We will write  (\ref{KG4}) as 
\begin{equation}\label{KG5} 
\ti\psi_+(x,\om)=\ti \al(\om)e^{ik(\om)|x|},\qquad \rIm\om>0, 
\end{equation} 
where $\al(t):=\psi_+(0,t)$. 
A nontrivial observation is that equality 
(\ref{KG5}) 
of analytic functions 
implies the similar identity for their restrictions to the real axis: 
\begin{equation}\label{KG6} 
\ti\psi_+(x,\om+i0)=\ti \al(\om+i0)e^{ik(\om+i0)|x|},\qquad \om\in\mathbb R, 
\end{equation} 
where $\ti\psi_+(\cdot, \om + i0) $ and $\ti \al(\om + i0) $ are the corresponding 
quasi-measures with values in $H^1(\mathbb R) $ and $\mC$, respectively. 
The problem is that the factor $M_x(\om):=e^{ik(\om+i0) |x|} $ is not smooth 
in $\om$ 
at the points $\om=\pm m$, and so identity (\ref{KG6}) requires a justification. 
 
\begin{lemma}\label{lcc} {\rm (\cite[Proposition 3.1]{KK2007})} 
For each  $x\in\mathbb R$, 
\begin{equation}\label{qm} 
\ti\al(\om+i\ve)\toQM\ti\al(\om+i0) \quad \mbox{\rm and} \quad G_x(\om+i\ve)\toL G_x(\om+i0)\qquad \mbox{\rm as}\quad \ve\to 0+, 
\end{equation} 
where $\ti G_x(\om+i\ve)=M_x(\om+i\ve)$ and $\ti G_x(\om+i0)=M_x(\om+i0)$. 
\end{lemma} 
 
Now (\ref{KG6}) follows from Lemma \ref{lQM}. 
\medskip 
 
Finally, the inversion of the Fourier transform can be  written as 
\begin{equation}\label{FLi} 
\psi_+(x,t)=\fr1{2\pi }\int_\mathbb R e^{-i\om t} \ti\psi_+(x,\om+i0)\,d\om 
=\fr1{2\pi }\int_\mathbb R e^{-i\om t} 
\ti \al(\om+i0)e^{ik(\om+i0)|x|}d\om, \qquad t>0,\quad x\in\mathbb R. 
\end{equation} 
 
 
\subsection{A nonlinear analogue of the Kato theorem} 
It turns out that properties of the quasi-measure $\ti\al(\om+i0)$ for $|\om|< m$ and for $|\om|> m$ 
differ greatly. This is due to the fact that the set $\{i\om: |\om|\ge m\}$ coincides, up to the factor $i$, 
with the continuous spectrum of the generator 
\begin{equation}\label{KGA} 
A=\left( 
\begin{array}{cc} 0&1\\ 
\fr{d^2}{dx^2}-m^2&0 
\end{array} 
\right) 
\end{equation} 
of the linear part of \eqref{KG1}. 
The following proposition plays the key role in our proofs.
It
is a~non-linear analogue of the Kato theorem 
on the absence of embedded eigenvalues in the continuous spectrum. 
Let us denote $\Si:=\{\om\in\mathbb R:|\om|> m\}$, and we will write below $\ti\al(\om)$ 
and  $k(\om)$
instead of $\ti\al(\om+i0)$ and  $k(\om+i0)$ for $\om\in\R$. 
 
\begin{pro}\label{lKato} {\rm (\cite[Proposition 3.2]{KK2007})} 
Let conditions  \eqref{C1}, \eqref{C2} and \eqref{C3} hold and let 
$\psi(t)$ be a~finite energy solution of equation \eqref{KG1}. Then the 
distribution $\ti\al(\om):=\ti \al (\om+i0)$ is absolutely continuous on $\Si$, and 
$\ti \al\in L^1(\Si)$. Moreover, 
\begin{equation}\label{Kato} 
\int_\Si |\ti \al(\om)|^2\5|\om\5 k(\om)|\5 d\om<\infty. 
\end{equation} 
\end{pro} 
The proof  \cite{KK2007} relies on the integral representation (\ref{FLi}), 
the \textit{a~priori estimate} \eqref{apri}, and uses some ideas of the Paley--Wiener theory. 
The main idea is that the functions $e^{ik(\om+i0)|x|}$ in 
(\ref{FLi})
do not belong to $H^1(\R)$ for $\om\in\Si$.

\subsection{Dispersive and bound components} 
Proposition \ref{lKato} suggests the splitting of 
the solution (\ref{FLi}) into the `dispersion' and `bound'  components 
\begin{eqnarray} 
\psi_+(x,t)&=&\fr1{2\pi }\int_\Si (1-\zeta(\om))e^{-i\om t} \ti \al(\om)e^{ik(\om)|x|}d\om+ 
\fr1{2\pi }\langle \ti \al(\om), \zeta(\om) e^{-i\om t}e^{ik(\om)|x|} \rangle 
\nonumber\\ 
\nonumber\\ 
&=&\psi_d(x,t)+\psi_b(x,t), \qquad t>0,\qquad x\in\mathbb R,  \label{FLi2} 
\end{eqnarray} 
where 
\begin{equation}\label{zeta} 
\zeta(\om)\in C_0^\infty(\mathbb R), \qquad \zeta(\om)=1 \quad \mbox{\rm for} \quad \om\in[-m-1,m+1], 
\end{equation} 
and 
$\langle\,\cdot{,}\,\cdot\rangle$ 
is the duality between quasi-measures and the corresponding test functions (in particular, Fourier transforms of functions from 
$L^1(\mathbb R)$). 
Note that $\psi_d(x,t)$ is a~dispersion wave, because 
\begin{equation}\label{psidRL} 
\psi_d(x,t):=\fr1{2\pi }\int_\Si (1-\zeta(\om))e^{-i\om t} \ti \al(\om)e^{ik(\om)|x|}d\om\xrightarrow
[t\to\infty]{} 0 
\end{equation} 
by \eqref{Kato} and the Lebesgue--Riemann theorem. 
The meaning of this convergence is specified in the following simple lemma.

\begin{lemma}\label{ld} {\rm (\cite[Lemma 3.3]{KK2007})} 
$\psi_d(x,t)$ 
is a~bounded continuous function of 
$t\in\mathbb R$ with values in $H^1(\mathbb R)$, and 
\begin{equation}\label{psid} 
(\psi_d(\cdot,t),\dot\psi_d(\cdot,t))\to 0 
\end{equation} 
in the seminorms \eqref{cER}. 
\end{lemma} 
Hence, it remains to prove the attraction 
 (\ref{gap}) for $Y_b(t):=(\psi_b(\cdot,t),\dot \psi_b(\cdot,t))$ instead of $Y(t)$: 
\begin{equation}\label{gapb} 
Y_b(t)\xrightarrow
[t\to +\infty]{} {\cal S}. 
\end{equation}

\subsection{Compactness and omega-limit trajectories} 
To prove (\ref{gapb}) we note, first, that the bound component $\psi_b(x,t) $ is a~smooth function, and 
\begin{equation}\label{gab2} 
\pa_x^j\pa_t^l \psi_b(x,t)=\fr1{2\pi }\langle \ti \al(\om), \zeta(\om)(ik(\om)\operatorname{sgn}\5 x)^j(-i\om)^l e^{-i\om t}e^{ik(\om)|x|} \rangle, 
 \qquad t>0,\quad x\in\mathbb R, 
\end{equation} 
which implies  the boundedness of each derivative: 
 
\begin{lemma}\label{lbound} {\rm (\cite[Proposition 4.1]{KK2007})} 
For any $j,l=0,1,2,\dotsc$ and $R>0$ 
\begin{equation}\label{gab22} 
\sup_{0<|x|\le R}\,\,\sup_{t\in \mathbb R}|\pa_x^j\pa_t^l \psi_b(x,t)|<\infty. 
\end{equation} 
\end{lemma} 
 
\textbf{Proof.} It suffices to 
verify that $\zeta(\om)k^j(\om)\om^l e^{-i\om t}e^{ik(\om)|x|}=\ti g_x(\om)$, where $g_x(\cdot)$ belongs 
to a bounded subset of $L^1(\mathbb R)$ for $0<|x|\le R$. 
Then (\ref{gab22}) follows from (\ref{gab2}) by the Parseval identity, inasmuch as 
$\al(t):=\psi(0,t)$ is a~bounded function. \bob
\medskip 
 
Hence, by the Ascoli--\allowbreak Arzela theorem, for any sequence 
 $s_j\to\infty$ there exists a~subsequence 
$s_{j'}\to\infty$, for which 
\begin{equation}\label{gab3} 
\pa_x^j\pa_t^l \psi_b(x,s_{j'}+t)\to \pa_x^j\pa_t^l \beta(x,t), \qquad (x,t)\in\mathbb R^2, 
\end{equation} 
the convergence being uniform on compact sets.  We will call any such function 
$\beta(x,t)$  an \textit{omega-limit trajectory} of the solution $\psi(x,t)$. It follows from bounds (\ref{gab22}) that 
\begin{equation}\label{betab} 
\sup_{(x,t)\in\mathbb R^2}|\pa_x^j\pa_t^l \beta(x,t)|<\infty. 
\end{equation} 
 
\begin{lemma}\label{lom} 
Attraction \eqref{gapb} is equivalent to the fact that 
any omega-limit trajectory is a~stationary orbit: 
\begin{equation}\label{gab4} 
\beta(x,t)=\phi_{\om_+}(x)e^{-i\om_+ t},\qquad \om_+\in\mathbb R. 
\end{equation} 
\end{lemma} 
This lemma follows from the uniform convergence (\ref{gab3}) on each compact set and the definition of the metric \eqref{metr}. 
 
\subsection{Spectral representation of omega-limit trajectories} 
Let us note that $\psi_b(x,t) $ is a~bounded function of 
$t \in \mathbb R $ with values in $H^1 (\mathbb R) $ due to the similar boundedness of $\psi_+(x,t)$ and $\psi_d (x,t)$. 
Therefore, $\psi_b (x,\cdot) $ is a~bounded function of $t\in \mathbb R^2 $ for each $x \in \mathbb R$, 
and convergence (\ref{gab3}) with $j = l = 0 $ implies 
the convergence of the corresponding Fourier transforms in time 
in the sense of 
tempered distributions. Moreover, this convergence holds in the sense of Ascoli--\allowbreak Arzela quasi-measures (\ref{QMA}) 
\begin{equation}\label{gab5} 
\ti\psi_b(x,\om)e^{-i\om s_{j'}}\toQM \ti\beta(x,\om),\qquad \forall x\in\mathbb R. 
\end{equation} 
Hence, representation (\ref{gab2}) implies that 
\begin{equation}\label{gab6} 
\zeta(\om)\ti\al(\om)e^{ik(\om)|x|} e^{-i\om s_{j'}} \toQM \ti\beta(x,\om),\qquad \forall x\in\mathbb R. 
\end{equation} 
Further, $e^{-ik(\om)|x|} $ is a~multiplier in the space 
of Ascoli--\allowbreak Arzela quasi-measures according \cite[Lemma B.3]{KK2007}). Now \eqref{gab6} gives that 
\begin{equation}\label{gab7} 
\zeta(\om)\ti\al(\om)e^{-i\om s_{j'}}  \toQM \ti\ga(\om):=\ti\beta(x,\om)e^{-ik(\om)|x|},\qquad \forall x\in\mathbb R. 
\end{equation} 
Hence, (\ref{gab2}) with $j = l = 0$ and $t + s_{j'}$ instead of $t$, 
gives in the limit $j'\to\infty$ 
the integral representation 
\begin{equation}\label{gab8} 
\beta(x,t)=\fr1{2\pi }\langle \ti\ga(\om)e^{ik(\om)|x|}, e^{-i\om t} \rangle,\qquad (x,t)\in\mathbb R^2, 
\end{equation} 
since $e^{ik(\om)|x|} $ is a~multiplier. Note that 
\begin{equation}\label{gab9} 
\beta(0,t)=\ga(t). 
\end{equation} 
Moreover, 
\begin{equation}\label{gab10} 
\supp\ti\ga\subset [-m,m] 
\end{equation} 
by \eqref{gab7} and Proposition \ref{lKato} due to the Riemann--Lebesgue theorem.

 
\subsection{Equation for omega-limit trajectories and spectral inclusion} 
Note that $\psi_+(x,t)$ is a~solution of \eqref{KG1} 
only for $t>0$ because of (\ref{gap2}) and (\ref{KG2}). However, the following simple but important lemma holds.

\begin{lemma}\label{leq} 
Any omega-limit trajectory satisfies the same equation \eqref{KG1}: 
\begin{equation}\label{KG11} 
\ddot\beta(x,t)=\beta''(x,t)-m^2\beta(x,t)+\delta(x) F(\beta(0,t)),\qquad (x,t)\in\mathbb R^2. 
\end{equation} 
\end{lemma} 
The lemma follows by substitution $\psi_+(x,s_{j'}+t)=\psi_d(x,s_{j'}+t)+\psi_b(x,s_{j'}+t)$ 
into equation (\ref{KG2}) and subsequent limit 
$s_{j'}\to\infty$ taking into account \eqref{psid} and \eqref{gab3}. 
\medskip 
 
The following proposition implies (\ref{gapb}) by Lemma \ref{lom}. 
 
\begin{pro}\label{pT} 
Under the hypotheses of Theorem~{\rm \ref{t5}} 
any omega-limit trajectory is a~stationary orbit of the form \eqref{gab4}. 
\end{pro} 
First, (\ref{KG11}) in the Fourier transform becomes the stationary equation 
\begin{equation}\label{Fap1} 
-\om^2\ti\beta(x,\om)=\ti\beta''(x,\om)-m^2\ti\beta(x,\om)+\delta(x) \ti f(\om),\qquad (x,\om)\in\mathbb R^2, 
\end{equation} 
where $f(t):=F(\beta(0,t))=F(\ga(t))$ by (\ref{gab9}). 
Further, 
(\ref{Fap}) gives that 
\begin{equation}\label{Fap2} 
f(t)=a(|\ga(t)|)\ga(t)=A(t)\ga(t),\qquad A(t)=a(|\ga(t)|), \qquad t\in\mathbb R. 
\end{equation} 
Hence, in the Fourier transform we obtain the convolution $\ti f=\ti A* \ti \ga $, 
which exists by (\ref{gab10}). Respectively, 
(\ref{Fap1}) reads 
\begin{equation}\label{Fap3} 
-\om^2\ti\beta(x,\om)=\ti\beta''(x,\om)-m^2\ti\beta(x,\om)+\delta(x) [\ti A*\ti\ga](\om),\qquad (x,\om)\in\mathbb R^2. 
\end{equation} 
This identity implies the key \textit{\bf spectral inclusion} 
\begin{equation}\label{si} 
\supp \ti A*\ti\ga\subset \supp\ti\ga, 
\end{equation} 
since $\supp\ti\beta(x,\cdot)\subset\supp\ti\ga$ and $\supp\ti\beta''(x,\cdot)\subset\supp\ti\ga$ by (\ref{gab8}). 
Using this inclusion, we will deduce below 
Proposition \ref{pT} applying the fundamental Titchmarsh convolution theorem of harmonic analysis. 
 
\subsection{The Titchmarsh convolution theorem} 
In 1926, Titchmarsh proved a~theorem on the distribution of zeros of entire functions \cite{Tit26}, \cite[p.119]{Lev96}, which 
implies, 
in particular, the following corollary 
\cite[Theorem 4.3.3]{Hor90}: 
 
\textbf{Theorem.} \textit{Let $f(\om)$ and $g(\om)$ be distributions of $\om\in\mathbb R$ with bounded supports. Then 
\begin{equation}\label{tt} 
[\supp\5 f\!*\!g]=[\supp f]+[\supp g], 
\end{equation} 
where $[X]$ denotes the \textit{convex hull} of a~subset $X\subset\mathbb R$.} 
\medskip 
 
Let us note that $\supp\ti\ga$ is bounded by (\ref{gab10}). Therefore, $\supp \ti A$ is also bounded, since 
$A(t):=a(|\ga(t)|)$ is a~polynomial of $|\ga(t)|^2$ by 
\eqref{C4}. 
Now the spectral inclusion (\ref{si}) implies by the Titchmarsh theorem that 
\begin{equation}\label{ttr} 
[\supp \ti A]+[\supp\ti\ga]\subset [\supp\ti\ga], 
\end{equation} 
which gives $[\supp \ti A]=\{0\}$. Furthermore 
$A(t):=a(|\ga(t)|)$ is a~bounded function by (\ref{betab}), because $\ga(t)=\beta(0,t)$. 
Hence, $\ti A(\om)=C\de(\om)$. Thus, 
\begin{equation}\label{const} 
a(|\ga(t)|)=C_1,\qquad t\in\mathbb R. 
\end{equation} 
Now the strict nonlinearity condition \eqref{C4} also gives that 
\begin{equation}\label{const2} 
|\ga(t)|=C_2,\qquad t\in\mathbb R. 
\end{equation} 
It is easy to deduce 
from this identity 
that $\supp\ti\ga = \{\om_ + \}$ by the same Titchmarsh theorem. 
Hence, $\ti\ga(\om)=C_3\5\de(\om-\om_+)$, which implies \eqref{gab4} by~\eqref{gab8}. 
 
 \begin{remark}\label{rST} 
{\rm
In the case of the Schr\"odinger equation \eqref{S1} the Titchmarsh theorem does not work. 
The point is that 
 the continuous spectrum of the operator $-d^2/dx^2 $ is the half-line $[0, \infty)$, 
so that the unbounded half-line $(-\infty, 0)$ now plays  
the role of the `spectral gap'. 
Respectively,  in this case inclusion \eqref{I} 
goes to $\supp\ti\beta (x, \cdot) \subset (-\infty,0)$, while the Titchmarsh theorem is applicable only to 
distributions with bounded supports.} 
\end{remark}

\subsection{Dispersion radiation and nonlinear energy transfer}\label{Sec-4.8} 
Let us give an informal comment on the proof of 
Theorem \ref{t5} behind  the formal arguments. 
The key part of the proof is concerned with the study of omega-limit trajectories of a~solution 
\begin{equation}\label{oltr} 
\beta(x,t)=\lim_{s_{j'}\to\infty} \psi(x,s_{j'}+t). 
\end{equation} 
First, Proposition \ref{lKato} implies the inclusion (\ref{gab10}), which gives 
\begin{equation}\label{I} 
\supp \ti\beta(x,\cdot)\subset [-m,m],\qquad x\in\mathbb R 
\end{equation} 
according to (\ref{gab8}). 
Next the Titchmarsh theorem allows us to conclude that 
\begin{equation}\label{II} 
\supp \ti\beta(x,\cdot)\subset \{\om_+\}. 
\end{equation} 
These two inclusions are suggested by the following informal ideas: 
\medskip\\ 
\textbf{A.} \textit{Dispersion radiation in the continuous spectrum.} 
\medskip\\ 
\textbf{B.} \textit{Nonlinear inflation of the spectrum and energy transfer.} 
\bigskip\\ 
\textbf{A. Dispersion radiation.} Inclusion (\ref{I}) is suggested by the dispersion mechanism, which is illustrated by 
energy radiation in a~wave field under harmonic excitation with frequency lying in the continuous spectrum. 
Namely, let us  consider the three-dimensional linear Klein--\allowbreak Gordon equation with the 
\textit{harmonic source} 
$$ 
\ds\ddot\psi(x,t)=\De\psi(x,t)-m^2\psi(x,t)+b(x)e^{i\om_0 t},\qquad x\in\mathbb R^3, 
$$ 
where $b\in L^2(\mathbb R^3)$. 
For this equation the \textit{limiting amplitude principle} holds  \cite{KopK2012-1, Lad1957, Mor1962}: 
\begin{equation}\label{lap} 
\psi(x,t)\sim a(x)e^{i\om_0 t},\qquad t\to\infty, 
\end{equation} 
where $a(x)$ is a~solution to the \textit{stationary Helmholtz equation} 
$$ 
-\om_0^2 a(x)=\De a(x)-m^2a(x)+b(x),\qquad x\in\mathbb R^3. 
$$ 
It turns out that the properties of the limiting amplitude $a(x)$ 
differ greatly for the cases 
$|\om_0|<m$ and $|\om_0|\ge m$. 
Namely, 
\begin{equation}\label{lapa} 
a(x)\in H^2(\mathbb R^3) \quad \mbox{for}\quad |\om_0|<m,\quad 
\mbox{but} \quad a(x)\not\in L^2(\mathbb R^3) \quad \mbox{for}\quad |\om_0|\ge m. 
\end{equation} 
This is obvious from the explicit formula in the Fourier transform 
\begin{equation}\label{lapaf} 
\hat a(k)=-\fr{\hat b(k)}{k^2+m^2-(\om+i0)^2},\qquad k\in\mathbb R^3. 
\end{equation} 
By (\ref{lap}) and (\ref{lapa}), 
the energy of the solution $\psi(x,t)$ tends to infinity for large times 
if $|\om_0|\ge m$. This means that the energy is transferred from the 
harmonic source to the wave field! 
In contrast, for $|\om_0 |<m$ 
the energy of the solution remains bounded, so that there is no radiation. 
 
Exactly this radiation in the case $|\om_0|\ge m$ 
prohibits the presence of harmonics with such frequencies in 
omega-limit trajectories, because the  
finite energy 
solution cannot radiate indefinitely. 
These arguments make natural the inclusion (\ref{I}), although its rigorous proof, as given above, 
is quite different.

\begin{figure}[htbp] 
\begin{center} 
\includegraphics[width=0.9\columnwidth]{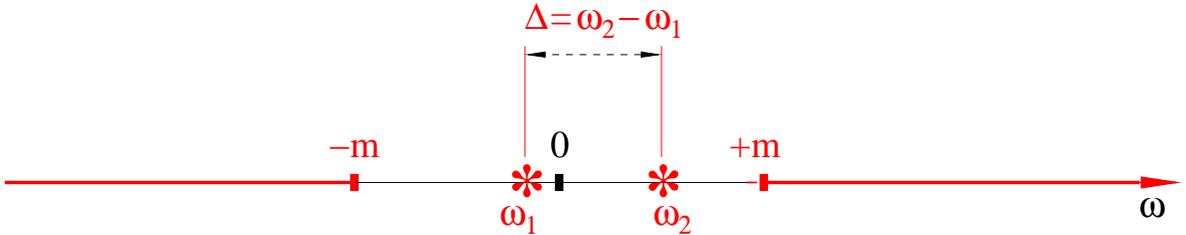} 
\caption{Two-point spectrum} 
\label{fig-11} 
\end{center} 
\end{figure}

Recall that the set $\Si:=\{\om\in\mathbb R$, $|\om|\ge m\}$ coincides with the continuous spectrum of the generator of the 
Klein--\allowbreak Gordon equation up to a~factor $i$. Note that the radiation in the continuous spectrum 
is well known in the theory of waveguides for a~long time. Namely, the waveguides 
 only pass signals with frequency greater than the threshold frequency, which is the edge point of 
continuous spectrum \cite{Lewin}. 
\smallskip\\
\textbf{B. Nonlinear inflation of spectrum and energy transfer.} For convenience, we will  call the \textit{spectrum of a~distribution} the support of its Fourier transform. 
Inclusion (\ref{II}) is due to an inflation of the spectrum by nonlinear functions. For example, let us consider the potential 
$U(|\psi|^2)\!=\!|\psi|^4\!$ and respectively, $F(\psi)=-\na_\psi U(|\psi|^2)= -4|\psi|^2\psi$. 
Consider the sum of two harmonics $\psi(t)=e^{i\om_1t}+e^{i\om_2t}$ whose spectrum is shown in Fig.~\ref{fig-11}, 
and substitute the sum 
into this nonlinearity. Then we obtain  
$$ 
F(\psi(t))\sim\psi(t)\ov{\psi(t)}\psi(t)=e^{i\om_2t}e^{-i\om_1t}e^{i\om_2t}+\dotsc 
=e^{i(\om_2+\De)t}+\dotsc \qquad \De:=\om_2-\om_1. 
$$ 
 
\begin{figure}[htbp] 
\begin{center} 
\includegraphics[width=0.9\columnwidth]{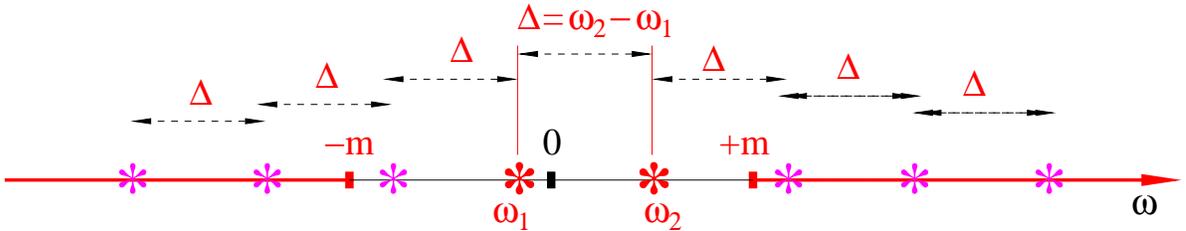} 
\caption{Nonlinear inflation of spectrum} 
\label{fig-2} 
\end{center} 
\end{figure}

The spectrum of this expression contains the harmonics with new frequencies $\om_1-\De$ and $\om_2 +\De $. 
As a~result, all the frequencies $\om_1-\De$, $\om_1-2\De , \dotsc$ and $\om_2 +\De $, $ \om_2 +2\De$, $ \dotsc$ 
will also appear in the dynamics 
(see Fig.~\ref{fig-2})). 
\bigskip

 
~\\
Therefore, the frequency lying in the continuous spectrum $|\om_0| \ge m$ will necessarily appear, 
causing the radiation of energy. This radiation will continue 
until the spectrum of the solution contains at least two different frequencies. 
Exactly
this fact 
prohibits the presence of two different frequencies in 
omega-limit trajectories, because the  
finite energy 
solution cannot radiate indefinitely. 
 
Let us emphasize  that 
the spectrum inflation by polynomials 
is established by the Titchmarsh convolution theorem, 
since the Fourier transform of a~product of functions equals the convolution of their Fourier transforms. 
 
\begin{remark}\label{phys} 
{\rm Physically the arguments above suggest the following  nonlinear radiation mechanism: 
\medskip\\
i) The nonlinearity inflates the spectrum which means the energy transfer from lower to higher modes;
\smallskip\\
ii) Then the dispersion radiation of the higher modes transports their energy
to infinity.
\smallskip\\
We have  justified  this radiation mechanism for the first time for the nonlinear 
$U(1)$-invariant
equations \eqref{KG1} and
\eqref{KGN}--\eqref{Dn}.
Our numerical experiments confirm the same radiation mechanism for nonlinear relativistically-invariant  
wave equations, see Remark \ref{rrm}.

} 
\end{remark}

\setcounter{equation}{0} 
 
\section{Global attraction to solitons} 
Here we describe the results of global attraction to solitons 
\eqref{att} for translation-invariant equations. 
 
\subsection{Translation-invariant wave-particle system} 
In \cite{KS1998}, we 
considered the system \eqref{w3}--\eqref{q3} with zero potential $V=0$: 
\begin{equation}\label{wq3} 
\left\{ 
\begin{array}{rlrl} 
\dot\psi(x,t)\!\!&\!\!=\pi(x,t),      &\dot\pi(x,t)\!\!&\!\!=\Delta\psi(x,t)-\rho(x-q(t)), \qquad x\in\mathbb R^3 
\\ 
\\ 
\dot q(t)\!\!&\!\!=\ds\frac{p(t)}{\sqrt{1+p^2(t)}}, &\dot p(t)\!\!&\!\!=-\ds\int\nabla\psi(x,t)\rho(x-q(t))\,dx. 
\end{array}\right| 
\end{equation} 
The corresponding Hamiltonian reads 
\begin{equation}\label{Ham0} 
\cH_0(\psi,\pi, q,p)=\ds\frac12\int[|\pi(x)|^2+|\na\psi(x)|^2]\,dx+\int\psi(x)\rho(x-q)\,dx+ 
\sqrt{1+p^2}, \end{equation} 
which coincides with (\ref{Ham}) for $V=0$. 
It is conserved along trajectories of the system \eqref{wq3}. Furthermore, 
this system is translation-invariant, 
and the corresponding total momentum  
\begin{equation}\label{P} 
P=p-\int \pi(x)\na\psi(x)\,dx. 
\end{equation} 
is also conserved. 
The system \eqref{wq3} admits traveling wave solutions (solitons) 
\begin{equation}\label{solit} 
\left\{ 
\begin{array}{rlrl} 
\psi_{v,a}(x,t)\!\!&\!\!=\psi_{v} (x-v t-a), &\pi_{v,a}(x,t)\!\!&\!\!=\pi_{v} (x-v t-a) 
\\ 
\\ 
q_{v,a}(t),\!\!&\!\!=vt+a &p_v\!\!&\!\!:=v/\sqrt{1-v^2} 
\end{array}\right| 
\end{equation} 
where $v,a\in\mathbb R^3 $ with $|v| <1 $. 
The set of these solitons form a~$6$-dimensional \textit{solitary submanifold} in $\cE$: 
\begin{equation}\label{cS} 
\cS=\{S_{v,a}=(\psi_{v} (x-a), \pi_{v} (x-a), a, p_v):\quad v,a\in\mathbb R^3,\quad |v|<1\} 
\end{equation} 
 
The main result of \cite{KS1998} is the following theorem. 
 
\begin{theorem}\label{t6} 
Let the Wiener condition \eqref{W1} hold. Then, for any finite energy solutions to the system \eqref{wq3}, 
\begin{equation}\label{dq} 
\dot q(t)\xrightarrow
[t\to\pm\infty]{} v_\pm. 
\end{equation} 
Moreover, for the field components the soliton asymptotics hold, 
\begin{equation}\label{ssol} 
(\psi(x,t), \pi(x,t)) 
=
(\psi_{v_\pm} (x-q(t)), \pi_{v_\pm} (x-q(t)))+
(r_\pm(x,t) ,s_\pm(x,t) )
\end{equation} 
where the remainders 
locally decay in the moving frame of the particle: for every $R> 0$ 
\begin{equation}\label{ssolh} 
\Vert\na r_\pm(q(t)+x,t)\Vert_R+\Vert r_\pm(q(t)+x,t)\Vert_R+\Vert s_\pm(q(t)+x,t)\Vert_R \xrightarrow
[t\to\pm\infty]{} 0. 
\end{equation} 
\end{theorem} 
The proof \cite{KS1998} relies on  a) 
the relaxation of acceleration (\ref{rel}) 
which holds for $V= 0$ (see Remark \ref{WC1} i)),
and 
b) on the \textit{canonical change of variables} to the comoving frame. 
The key role plays the fact that the soliton $S_{v,a} $ minimizes the Hamiltonian (\ref{Ham0}) 
under fixed total momentum \eqref{P}, implying the \textit{orbital stability of solitons} \cite{GSS87, GSS90}. 
Furthermore, the 
\textit{strong Huygens principle} for the 3D wave equation is used. 
  
  \begin{remark}
{\rm 
The Wiener condition~\eqref{W1} is sufficient for the relaxation~\eqref{rel}
of solutions to translation-invariant system ~\eqref{wq3}.
However it is not  
necessary:
 for example,~\eqref{rel} obviously holds for $\rho(x)\equiv 0$.
 Moreover, \eqref{rel}
 holds also in the case of 
 small $\Vert\rho\Vert$, see Section 4.3. }

 \end{remark}

\subsection{Translation-invariant  Maxwell-Lorentz  equations} 
 
In \cite{IKM2004}, asymptotics of type (\ref{dq})--(\ref{ssolh}) 
were extended to the translation-invariant Maxwell--Lorentz system \eqref{ML} 
with zero external fields. In this case, the Hamiltonian \eqref{HAM}  reads as 
\begin{equation}\label{HAM0} 
\cH_0=\frac12\int[E^2(x)+B^2(x)]\,dx+\sqrt{1+p^2}. 
\end{equation} 
The extension of  the arguments \cite{KS1998} to this case required an essential 
analysis of the corresponding Hamiltonian structure which is necessary for the canonical transformation.
Now the key role 
in application of the strong Huygens principle
play novel estimates for the decay 
of oscillations of the  Hamiltonian 
(\ref{HAM0}) and of total momentum 
along solutions to a perturbed Maxwell-Lorentz system, see \cite[(4.24) and (4.25)]{IKM2004}.

 \subsection{Weak coupling} 
Asymptotics of type (\ref{dq})--(\ref{ssolh}) in a~stronger form were proved for the system \eqref{w3}--\eqref{q3} 
under the weak coupling condition 
\begin{equation}\label{rosm} 
\Vert\rho\Vert_{L^2(\mathbb R^3)}\ll 1. 
\end{equation} 
Namely, 
in \cite{IKS2004a} 
we have considered initial fields with a decay $|x|^{-5/2-\si}$ with a parameter $\si>0$
(condition (2.2) of \cite{IKS2004a}),  
assuming that
\begin{equation}\label{nV0} 
 \na V(q)= 0,\qquad\qquad |q|>\co. 
\end{equation}
Under these assumptions we prove
the strong relaxation 
\begin{equation}\label{dqsm} 
|\ddot q(t)|\le C(1+|t|)^{-1-\si},\qquad t\in\R
\end{equation} 
for "outgoing" solutions which satisfy the condition 
\begin{equation}\label{sfin} 
|q(t)|\xrightarrow
[t\to\pm\infty]{}\infty.
\end{equation} 
In particular, all solutions are outgoing in the case $V(q)\equiv0$.
Asymptotics 
(\ref{dq})--(\ref{ssolh}) 
under these assumptions are refined 
similarly to (\ref{at4}): 
\begin{equation}\label{at4sol} 
\dot q(t)\to v_\pm,~~ (\psi(x,t), \pi(x,t)) 
\!=\! 
(\psi_{v_\pm} (x\! -\! q(t)), \pi_{v_\pm} (x\! -\! q(t)))+W(t)\Phi_\pm+(r_\pm(x,t) ,s_\pm(x,t)),~~ t\to\pm\infty. 
\end{equation} 
Here the `dispersion waves' 
$~W(t)\Phi_\pm$ are solutions to the free wave equation, and the remainder 
now converges to zero in the global energy norm: 
 \begin{equation}\label{ssolhsm} 
\Vert\na r_\pm(q(t)+x,t)\Vert+\Vert r_\pm(q(t)+x,t)\Vert+\Vert s_\pm(q(t)+x,t)\Vert \xrightarrow
[t\to\pm\infty]{} 0. 
\end{equation}

\begin{remark}\label{rdr}
{\rm 
This progress with respect to the local decay \eqref{ssolh} is due to the fact that we 
identify the dispersion wave $W(t)\Phi_\pm$
under the smallness condition \eqref{rosm}. This identification is possible by the decay
rate \eqref{dqsm} which is more strong than \eqref{rel}.}
\end{remark}

\begin{figure}[htbp] 
\begin{center} 
\includegraphics[width=0.8\columnwidth]{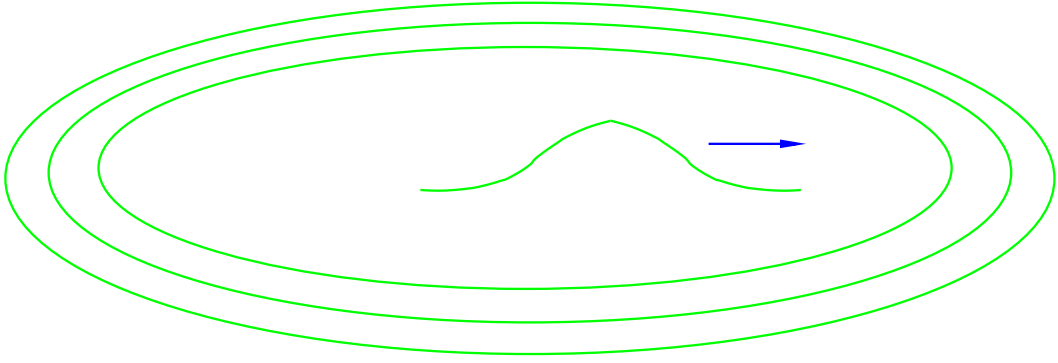} 
\caption{Soliton and dispersion waves} 
\label{fig-32} 
\end{center} 
\end{figure}

The solitons propagate with velocities less than $1$, and therefore 
they separate at large time from the dispersion waves $W(t) \Phi_\pm $, which propagate with unit velocity (Fig.~\ref{fig-32}). 
 
 The proofs rely on the integral Duhamel  representation and rapid  dispersion decay for the free wave equation.
A similar result was obtained in \cite{IKM2003} for a~system of type \eqref{w3}--\eqref{q3} with the Klein--\allowbreak Gordon equation, 
and in \cite{IKS2002}, for the system \eqref{ML} under the same condition (\ref{sfin})
assuming that $E^{\rm ext}(x)= B^{\rm ext}(x)= 0$  for $|x|>\co$. 
In \cite{IKS2004b}, 
this result was extended to a~system of type \eqref{ML} with a~rotating charge in the Maxwell field. 

\begin{remark}\label{rGC}
{\rm 
The results \cite{IKS2004a}--\cite{IKS2004b} imply 
the 
"Grand Conjecture" \cite[p.460]{soffer2006}
in the moving frame 
for the corresponding systems with $V(q)\equiv 0$ and  $E^{\rm ext}(x)\equiv B^{\rm ext}(x)\equiv 0$  
under the smallness condition \eqref{rosm}.
}
\end{remark}

\bigskip
 
\subsection{Solitons of relativistically-invariant equations} 
The existence of soliton solutions $\psi(x-vt)$ was extensively studied in the 1960--1980's 
for a~wide class of relativistically-invariant $U(1)$-invariant nonlinear wave equations 
\begin{equation}\label{NWEn} 
\ddot \psi (x,t) = 
\De\psi(x,t) + F(\psi(x,t)), \qquad x \in \mathbb R^n. 
\end{equation} 
Here $F(\psi)=-\na_{\ov\psi} U(\psi)$, where $U(\psi)=u(|\psi|)$ with 
$u\in C^2(\mathbb R)$. 
In this case, equation (\ref{NWEn}) is equivalent to the Hamilton system of type (\ref{w1ham}) with a~conserved in time 
Hamilton functional 
\begin{equation}\label{HAMbpn} 
\cH(\psi,\pi)=\int [\frac12 |\pi(x)|^2+ \frac12 |\na\psi (x)|^2+ U(\psi(x)) ]\,dx. 
\end{equation} 
This equation is translation-invariant, 
so the total momentum 
\begin{equation}\label{momn} 
P:=-\int \pi(x)\na \psi(x)\,dx 
\end{equation} 
is also conserved. 
Furthermore, this equation is also $U(1)$-invariant; i.e., 
$F(e^{-i\theta}\psi)\equiv e^{i\theta}F(\psi)$ 
for 
$\theta \in [0,2 \pi]$. Respectively, it can admit 
 soliton solutions of the form $e^{-i\om t} \phi_\om(x)$. Substitution into (\ref{NWEn}) gives the nonlinear eigenfunction problem 
\begin{equation}\label{NEPn} 
-\om^2 \phi_\om(x)=\De\phi_\om(x) 
+F(\phi_\om(x)),\qquad x\in\mathbb R. 
\end{equation} 
Under suitable conditions on the potential $U$, 
solutions $\phi_\om\in H^1(\mathbb R^n)$ exist 
and decay exponentially as $|x|\to\infty$ 
for $\om\in\cO$, where $\cO$ is an open subset 
 of $\mathbb R$. 
 
The most general results on the existence of the solitons were obtained by Strauss, Berestycki and P.-L. Lions 
\cite{St77, BL83-1, BL83-2}. 
The approach \cite{BL83-2} relies on 
variational and topological methods of the Ljusternik--Schnirelman theory \cite{LS1934,LS1947}. 
The development of this approach in 
\cite{EGS} provided the existence of solitons for nonlinear relativistically-invariant 
Maxwell--\allowbreak Dirac equations (\ref{DM}).

The orbital stability of solitons has been studied by 
Grillakis, Shatah, Strauss, and others \cite{GSS87, GSS90}. However, the global attraction
to solitons (\ref{at14}) is still open problem.
 
\smallskip 
 
The equation (\ref{NWEn}) is also Lorentz-invariant. Hence, the solitons with any velocities $|v|<1$ are obtained from 
the `standing soliton' 
$e^{-i\om t} \phi_\om(x)$ via the Lorentz transformation 
\begin{equation}\label{swen} 
\psi_{v,\om}(x,t):=e^{-i\om \ga_v(t-vx)}\phi_\om(\ga_v(x-vt)),\qquad \ga_v:=\sqrt{1-v^2}. 
\end{equation} 
The total energy (\ref{HAMbpn}) and the total momentum (\ref{momn}) 
of the soliton coincide with the 
corresponding formulas for a~relativistic particle (see \cite[(4.1)]{edks}): 
\begin{equation}\label{Hswen} 
E_{v,\om}= 
\ds\fr{m_0(\om)}{\sqrt{1-v^2}},\qquad P_{v,\om}= 
\ds\fr{m_0(\om)v}{\sqrt{1-v^2}}, 
\end{equation} 
where $m_0(\om)>0$ for $\om\ne 0 $, provided \eqref{C2} holds. 
Therefore, the relativistic `dispersion relation' holds, 
\begin{equation}\label{Hdispn} 
E_{v,\om}^2=m_0^2(\om)+P_{v,\om}^2, 
\end{equation} 
which implies the Einstein's famous formula $E=m_0c^2$ if $v=0$ 
(recall that we set $c=1$). 
 \medskip
 
 In the one-dimensional case $n = 1$, equation (\ref{NEPn}) reads 
\begin{equation}\label{NEP} 
-\om^2 \phi_\om(x)=\phi_\om''(x) 
+F(\phi_\om(x)),\qquad x\in\mathbb R. 
\end{equation} 
This ordinary differential equation is easily solved in quadratures using the `energy integral' 
\begin{equation}\label{NEPen} 
\fr12 |\phi_\om'(x)|^2-U(\phi_\om(x))+\fr12 \om^2 |\phi_\om(x)|^2=\co,\qquad x\in\mathbb R. 
\end{equation}

\begin{figure}[htbp] 
\begin{center} 
\includegraphics[width=0.8\columnwidth]{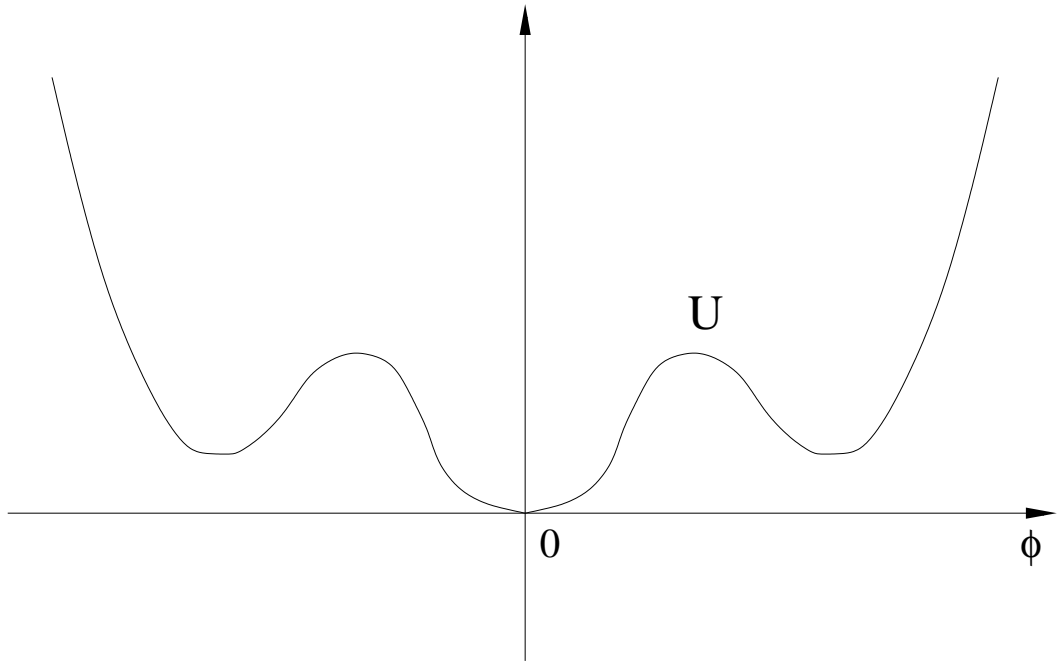} 
\caption{The potential $U$} 
\label{fig-5} 
\end{center} 
\end{figure} 
\begin{figure}[htbp] 
\begin{center} 
\includegraphics[width=0.8\columnwidth]{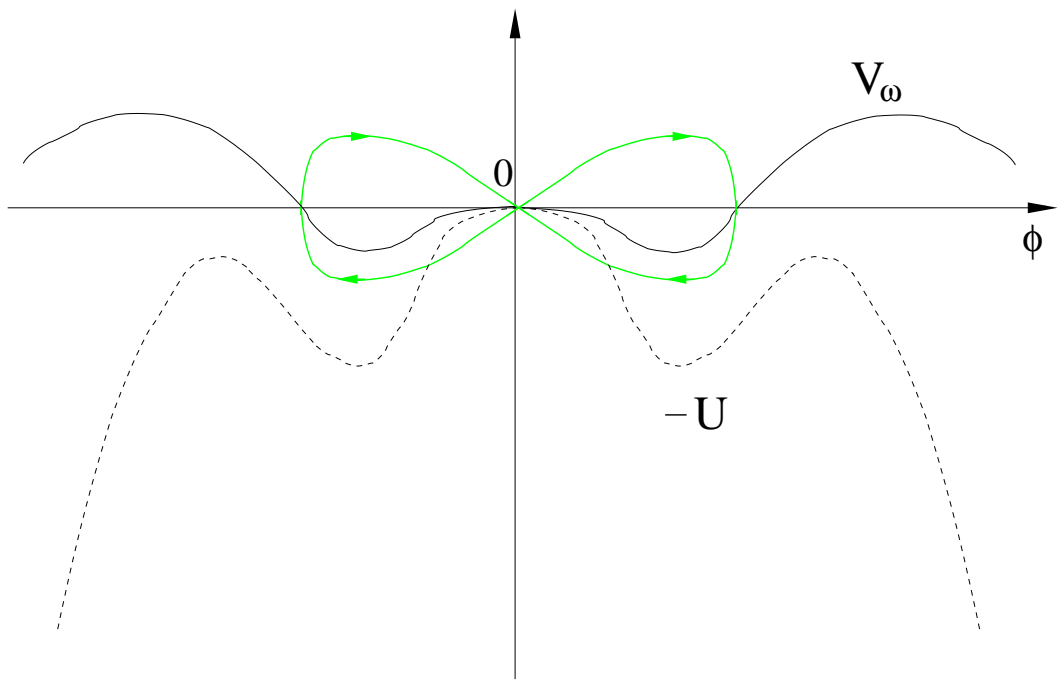} 
\caption{Potentials and soliton on the phase plane} 
\label{fig-6} 
\end{center} 
\end{figure} 
 
This identity shows that finite energy solutions to the equation (\ref{NEPen}) 
exist for potentials $U$, similar to shown in Fig.~\ref{fig-5}. 
Namely, the potential $V_\om(\phi):=-U(\phi)+\fr12 \om^2 |\phi|^2$ with $\om^2< U''(0)$ 
has the shape represented in 
Fig.~\ref{fig-6}, guarantying the existence of an exponentially decaying trajectory as $x\to\pm\infty$ 
(the green contour) 
which represents the soliton. 
 
\setcounter{equation}{0} 
 
\section{Adiabatic effective dynamics of solitons} 
Existence of solitons and soliton-type asymptotics (\ref{ssol}) are typical features of translation-invariant systems. 
However, if a~deviation of a~system from translation invariance is small in some sense, then the system may admit solutions 
that are permanently close to solitons with parameters depending on time (velocity, etc.). Moreover, in some cases it 
turns out possible to find 
an `effective dynamics' describing the evolution of these parameters.

\subsection{Wave-particle system with slowly varying external potential} 
Solitons (\ref{solit}) are solutions to the system 
\eqref{wq3} with zero external potential. 
However, even for the corresponding system
\eqref{w3}--\eqref{q3} with
a~nonzero external potential the {\it soliton-like} solutions of the form 
\begin{equation}\label{asol} 
\psi(x,t)\approx \psi_{v(t)}(x-q(t)) 
\end{equation} 
may exist
if the potential is slowly varying: 
\begin{equation}\label{asolV} 
|\na V(q)|\le \ve\ll 1. 
\end{equation} 
Now the total momentum (\ref{P}) is not conserved, but its slow evolution 
together with evolution of solutions (\ref{asol}) 
can be described 
in terms of finite-dimensional Hamiltonian dynamics. 
 
Let us denote by $P=P_v$ the total momentum of the soliton $S_{v,Q}$ 
in the notations (\ref{cS}),
and observe that 
the mapping $\cP: v\mapsto P_v$ is an isomorphism of the ball $|v| <1 $ onto $R^3$. 
Therefore, we can regard $Q,P$ as the global coordinates 
on the solitary manifold $\cS$ and define an effective Hamilton functional 
\begin{equation}\label{Heff} 
H_{\rm eff}(Q,P_v)\equiv \cH_0(S_{v,Q}),\qquad (Q,P_v)\in\cS,
\end{equation} 
where $ \cH_0$ is the {\it unperturbed} Hamiltonian  (\ref{Ham0}).
It is easy to observe that the functional admits the splitting 
$H_{\rm eff}(Q,\Pi)=E(\Pi)+V(Q)$, 
so that the corresponding Hamilton equations read 
\begin{equation}\label{dyn} 
\dot Q(t)=\na E(\Pi(t)),\qquad \dot \Pi(t)=-\na V( Q(t)). 
\end{equation} 
 
The main result of \cite{KKS1999} is the following theorem. 
 
\begin{theorem}\label{t7} 
Let condition (\ref{asolV}) hold,  
and let the initial state $(\psi_0,\pi_0,q_0,p_0)$ 
be a~soliton $S_0\in\cS$ with total momentum $P_0$. 
Then the corresponding solution 
$(\psi(x,t),\pi(x,t),q(t),p(t))$
to the system \eqref{w3}--\eqref{q3}  admits the following `adiabatic asymptotics' 
\begin{eqnarray} 
&&|q(t)-Q(t)|\le C_0, \quad |P(t)-\Pi(t)|\le C_1\ve \quad \mbox{\rm for}\quad |t|\le C\ve^{-1},\label{effd1} 
\\ 
\nonumber\\ 
&&\sup_{t\in\mathbb R} \Big[\Vert \na[\psi(q(t)+x,t)-\psi_{v(t)}(x)]\Vert_R+ \Vert \pi(q(t)+x,t)-\pi_{v(t)}(x)]\Vert_R\Big]\le C\ve, 
\label{effd2} 
\end{eqnarray} 
where 
$P(t)$ is the total momentum \eqref{P},
the velocity 
$v(t)=\cP^{-1}(\Pi(t))$, and 
$(Q(t),\Pi(t))$ is the solution to the effective Hamilton equations \eqref{dyn} with initial conditions 
\begin{equation}\label{inid} 
Q(0)=q(0),\qquad \Pi(0)=P(0). 
\end{equation} 
\end{theorem} 
Note that the relevance of effective dynamics (\ref{dyn}) is due to consistency of the Hamilton structures: 
\medskip\\ 
1) The effective Hamiltonian (\ref{Heff}) is the restriction of the Hamiltonian (\ref{Ham0}) 
onto the solitary manifold~$\cS$. 
\medskip\\ 
2) As shown in \cite{KKS1999}, the canonical form of the Hamilton system 
(\ref{dyn}) is also the restriction of the canonical form of the original system \eqref{w3}--\eqref{q3} onto $\cS$: 
\begin{equation}\label{can} 
P\,dQ=\Bigl[p\,dq+\int\psi(x) \,d\pi(x)\5 dx\Bigr]\Big|_\cS. 
\end{equation} 
Hence, the
total momentum $P$ is 
canonically conjugate to the variable $Q$ on the solitary manifold $\cS$.
This fact clarifies 
 definition  (\ref{Heff}) of the  effective Hamilton functional 
as the function of the  
total momentum $P_v$, 
rather than of the particle momentum $p_v$. 
\medskip 
 
One of main  results of \cite{KKS1999} is 
the following `effective dispersion relation': 
\begin{equation}\label{EP} 
E(\Pi)\sim \fr{\Pi^2}{2(1+m_e)}+{\rm const},\qquad |\Pi|\ll 1. 
\end{equation} 
It means that the non-relativistic mass of the 
slow soliton increases due to the interaction with the field by the value 
\begin{equation}\label{me} 
m_e=-\fr 13 \langle\rho,\De^{-1}\rho\rangle.
\end{equation} 
This increment 
is proportional to the field-energy of the soliton at rest, that
agrees with the Einstein principle of the mass-energy equivalence (see below).

\begin{remark}\label{rad}
{\rm
The relation (\ref{EP}) suggests only a hint that $m_e$ is the increment of the effective mass.
The genuine justification is given by  relevance of the adiabatic effective dynamics 
(\ref{dyn}) which is confirmed by the  asymptotics 
(\ref{effd1})--(\ref{effd2}).
}

\end{remark}

\subsection{Generalizations and the mass-energy equivalence} 
In \cite{KS2000ad}, asymptotics (\ref{effd1}), (\ref{effd2}) were extended to solitons of the Maxwell--Lorentz 
equations \eqref{ML} with small external fields, 
and the increment 
of the non-relativistic mass of type (\ref{me}) was calculated. 
It also turns out to be proportional to the own field energy of the static soliton. 
 
Such an equivalence of the own electromagnetic field energy of the particle and of its 
 mass was first suggested in 1902 by Abraham: he obtained by a~direct calculation that 
the electromagnetic self-energy $E_{\rm own}$ 
of the electron at rest contributes the increment $m_e=\ds\frac43 E_{\rm own} / c^2 $ into its nonrelativistic mass 
(see \cite{A1902, A1905}, and also \cite[pp.~216--217]{K2013}). 
It is easy to see that this self-energy is infinite for the point electron with the charge density $\de(x-q)$, because 
in this instance the Coulomb electrostatic field $|E(x)| \sim C / |x-q|^2$ as $x \to q$, so that the integral in (\ref{HAM}) diverges. 
Respectively, the field mass for a~point electron is infinite, which contradicts the experiment. This is why 
Abraham introduced the model of `extended electron' for which the self-energy is finite.

At that time Abraham put forth the idea 
that the whole mass of an electron is due to its own electromagnetic energy; i.e., $m = m_e$: `... \textit{the matter has disappeared, 
only the radiation remains}...', as wrote philosophically minded contemporaries \cite[pp.~63, 87, 88]{H1908} (Smile :)\,)

This idea was refined and developed in 1905 by Einstein, who has discovered the famous universal 
relation $E=m_0 c^2$ suggested by the relativity theory \cite{Einstein-1905}.
The extra factor $\frac43$ in the Abraham formula 
is due to the non-relativistic nature of the system \eqref{ML}. 
According to the modern view, about 80 \% of the electron mass has electromagnetic origin~\cite{F1966}.

Further, the asymptotics of type (\ref{effd1}), (\ref{effd2}) 
were obtained in \cite{FTY2002, FGJS2004} for the nonlinear Hartree 
and Schr\"odinger equations with slowly varying external potentials, and in \cite{S2010}--\cite{LS2009}, for nonlinear 
Einstein--\allowbreak Dirac, Chern--Simon--Schr\"odinger and Klein--\allowbreak Gordon-Maxwell equations
 with small external fields. 
 
Recently, a~similar adiabatic effective dynamics was established in \cite{BCFJS} for an electron in the second-quantized Maxwell field 
in presence of a~slowly varying external potential. 

\begin{remark}\label{rad2}
{\rm
The dispersion relation (\ref{Hdispn}) for relativistic solitons
   formally implies the 
Einstein's formula $E=m_0c^2$ if $v=0$ 
(recall that we set $c=1$).
However, its genuine dynamical justification requires the relevance of the corresponding 
adiabatic effective dynamics 
for the solitons 
with the relativistic kinetic energy $E=\sqrt{m_0^2+P^2}$. 
The first result of this type for relativistically-invariant  Klein--\allowbreak Gordon-Maxwell equations 
is established  in \cite{LS2009}.

}
\end{remark}

\medskip

\setcounter{equation}{0} 
\section{Asymptotic stability of solitary waves} 
 
The asymptotic stability 
of solitary manifolds means the local attraction; i.e., for the state sufficiently close to the manifold. 
The main peculiarity of this attraction is the instability of the dynamics \textit{along the manifold}. 
This follows directly from the fact that the solitary waves   
move with different velocities, and therefore run away over a~long time. 

Analytically, this instability is related to the presence of the discrete spectrum of 
the linearized dynamics with $\rRe\lam\ge 0$.
Namely, the tangent vectors to the 
solitary manifolds are the eigenvectors and the associated eigenvectors of the 
generator of the linearized dynamics at the solitary wave. They 
correspond to the zero eigenvalue. Respectively, 
the Lyapunov theory is not applicable in this case.

In a~series of papers an ingenious 
 strategy was developed for proving the asymptotic stability 
of solitary manifolds. 
 In particular, 
 this strategy 
 includes the 
symplectic projection of the trajectory onto the solitary manifold, 
the modulation equations for the soliton parameters of the projection, 
and the decay of the transversal component. 
This approach 
 is a~far-reaching development of the Lyapunov stability theory.

 \subsection{Linearization and decomposition of the dynamics}

The strategy was initiated in the pioneering works of Soffer and Weinstein 
\cite{W1985, SW1990, SW1992}; see the survey \cite{soffer2006}. 
 The results  
concern the nonlinear $U(1)$-invariant Schrodinger equation with a~real potential $V(x)$ 
\begin{equation}\label{Su} 
i\dot\psi(x,t)=-\Delta\psi(x,t)+V(x)\psi(x,t)+\lambda|\psi(x,t)|^p\psi(x,t),\qquad x\in\mathbb R^n, 
\end{equation} 
where $\lambda\in\mathbb R$, $p=3$ or $4$, $n=2$ or $n=3$, 
and $\psi(x,t)\in\mC$. 
The corresponding Hamilton functional reads 
\begin{equation}\label{hams} 
H=\int [\fr12|\na\psi|^2+\fr12V(x)|\psi(x)|^2+\fr\lambda p|\psi(x)|^p]\,dx. 
\end{equation} 
For $\lambda=0$ the equation (\ref{Su}) is linear. Let $\phi_*(x)$ denote its ground state corresponding to the minimal eigenvalue 
$\om_*<0$. Then  $ C\phi_*(x)e^{-i\omega_*t}$ 
are periodic solutions for any complex constant $C$. The corresponding phase curves are the circles filling the 
complex line (which is the real plane). 
 For nonlinear equations (\ref{Su}) with small real $\lam\ne 0$, it turns out that a~remarkable {\it bifurcation} occurs: 
a~small neighborhood of zero of the complex line 
is transformed into an analytic-invariant solitary manifold 
$\cS$
which is still filled by the circles 
 $\psi_\om(x)e^{-i\omega t}$ 
with frequencies $\om$ close to $\om_*$. 

The main result of \cite{SW1990, SW1992} (see also \cite{PW1997}) is the long time 
attraction to one of these trajectories 
at large times for any solution with sufficiently small initial data 
\begin{equation}\label{soli} 
\psi(x,t)=\psi_{\pm}(x)e^{-i\omega_{\pm}t}+r_\pm(x,t), 
\end{equation} 
where the remainder decays in the weighted norms: for $\si>2$ 
\begin{equation}\label{solir} 
\Vert \langle x\rangle^{-\si} r_\pm(\cdot,t)\Vert_{L^2(\mathbb R^n)} \xrightarrow
[t\to\pm\infty]{} 0, 
\end{equation} 
where $\langle x\rangle:=(1+|x|)^{1/2}$. 
The proofs rely on linearization of the dynamics, the  decomposition 
$$
\psi(t)=e^{-i\Theta(t)}(\psi_{\om(t)}+\phi(t)),
$$
and the orthogonality condition
\begin{equation}\label{or}
\langle \psi_{\om(0)}, \phi(t)\rangle=0
\end{equation} 
(see \cite[(3.2) and (3.4)]{SW1990}). This orthogonality 
and the dynamics (\ref{Su})
imply the 
{\it modulation equations} for $\om(t)$ and $\ga(t)$ where 
$\ga(t):=\Theta(t)-\ds\int_0^t\om(s)ds$ (see (3.2) and (3.9a), (3.9b) of
\cite{SW1990}.
The orthogonality 
(\ref{or}) ensures that $\phi(t)$ lies in the continuous spectral space of the Schr\"odinger operator 
$H(\om_0):=-\De+V+\lam|\psi_{\om_0}|^{m-1}$ which results in the time decay 
\cite[(4.2a) and (4.2b)]{SW1990}
of the component $\phi(t)$.
Finally, this decay implies 
the convergence $\om(t)\to\om_\pm$ and
the asymptotics (\ref{soli}) as $t\to\pm\infty$.
\smallskip 

  These results and methods were further developed by many authors 
for nonlinear Schr\"odinger, wave 
and Klein--\allowbreak Gordon equations with external potentials  
under various types of spectral assumptions on the linearized dynamics \cite{PW1997} - \cite{T2003}
for the case of small inital data. 
\medskip

A significant progress in this theory 
has been achieved by  Buslaev, Perelman and Sulem 
who have established
in \cite{BP1993}--\cite{BS2003}
the asymptotics of type 
(\ref{soli}) 
for the first time for
translation-invariant 1D Schr\"odinger equations 
\begin{equation}\label{BPS} 
i\dot\psi(x,t)=-\psi''(x,t)+F(\psi(x,t)),\qquad x\in\mathbb R
\end{equation} 
which are also $U(1)$-invariant. The latter means that the nonlinear function $F(\psi)$ satisfies
the identities 
(\ref{C3})--(\ref{U1}).
Then the corresponding solitons have the form 
$\psi(x,t)=\psi_{v,\om}(x-vt-a)e^{-i(\om t+\theta)}$.
The set of all solitons form 4-dimensional smooth 
submanifold $\cS$ 
of the Hilbert phase space $\cX:=L^2(\R)$.

The novel approach \cite{BP1993}--\cite{BS2003}
relies on the {\it symplectic projection} $P$ of solutions onto the solitary manifold.
This means that for $S:=P\psi$ we have
\begin{equation}\label{PS}
Z:=\psi-S\quad\mbox{is symplectic-orthogonal to the tangent space}\quad \cT:=T_{S}\cS.
\end{equation} 
The projection is well defined 
in a small neighborhood of $\cS$:
it is important that $\cS$ is the {\it symplectic manifold}, i.e. the symplectic form 
is nondegenerate on the tangent spaces  $T_{S}\cS$.
Now the  solution is decomposed 
into the {\it symplectic orthogonal}
components $\psi(t)=S(t)+Z(t)$ 
where $S(t):=P\psi(t)$, and 
the
dynamics is linearized at the solitary wave $S(t):=P\psi(t)$ for every $t>0$.
In particular, 
the approach \cite{BP1993}--\cite{BS2003} allowed to get rid of the smallness assumption 
on initial data.

The main results 
of \cite{BP1993}--\cite{BS2003}
are the asymptotics of type  (\ref{at4sol}), (\ref{soli})
for solutions with initial data close to the solitary manifold $\cS$: 
\begin{equation}\label{sdw} 
\psi(x,t)=\psi_\pm(x-v_\pm t)e^{-i\om_\pm t}+W(t)\Phi_\pm+r_\pm(x,t), 
\end{equation} 
where $W(t)$ is the dynamical group of the free Schr\"odinger equation, $\Phi_{\pm}$ are some finite energy states, and 
$r_{\pm}$ are the remainders which tend to zero in the global norm: 
\begin{equation}\label{glob2u} 
\Vert r_{\pm}(\cdot,t)\Vert_{L^2(\mathbb R)} \xrightarrow
[t\to\pm\infty]{} 0. 
\end{equation} 
The asymptotics are obtained under the condition 
\cite[(1.0.12)]{BS2003}
which means the 
 strong coupling of the discrete and continuous spectral components.
 This condition is the nonlinear version of 
the Fermi Golden Rule \cite{RS4} which 
 was originally introduced by Sigal \cite{Sig1993,MSig1999}.
In \cite{C2001}, 
these results were extended to \textit{n}D translation-invariant Schr\"odinger equations in dimensions $n\ge 2$. 
\medskip

 \subsection{Method of symplectic projection in the Hilbert space}
The proofs of asymptotics (\ref{sdw})--(\ref{glob2u}) in \cite{BP1993}--\cite{BS2003} rely on the 
linearization of the dynamics (\ref{BPS}) at the 
soliton $S(t):=P\psi(t)$ which is the  nonlinear  symplectic projection of  
$\psi(t)$ onto the solitary manifold $\cS$.
The Hilbert phase space 
$\cX:=L^2(\R)$
admits the splitting $\cX= \cT(t)\oplus\cZ(t)$, where 
${\cal Z}(t)$ is the
{\bf symplectic orthogonal} space to the tangent space $\cT(t):=T_{S(t)}\cS$. 
The corresponding equation for the {\it transversal component}  $Z(t)$ reads 
\begin{equation}\label{Zt} 
\dot Z(t)=A(t)Z(t)+N(t),
\end{equation} 
where $A(t)Z(t)$ is the linear part while $N(t)=\cO(\Vert Z(t)\Vert^2)$ is the corresponding 
nonlinear part. 
The main peculiarity of this equation is that it is {\it nonautonomous},
and the generators  $A(t)$ are nonselfadjoint (see Appendix \cite{KKsp2013}).
The main issue is that  $A(t)$
are {\it Hamiltonian operators}.
The strategy of \cite{BP1993}--\cite{BS2003} relies on the following ideas.
\medskip\\
{\bf S1. Modulation equations.} 
The parameters of the soliton $S(t)$ satisfy {\bf modulation equations}: for example,
for its velocity we have
$\dot v(t)=M(\psi(t))$, where $M(\psi)=\cO(\Vert Z\Vert^2)$ for 
small $\Vert Z\Vert$. Hence, the parameters vary extra slowly near the solitary manifold, 
like adiabatic invariants. 
\smallskip\\
{\bf S2. Tangent and transversal components.}
The transversal component
$Z(t)$
in the splitting $\psi(t)=S(t)+Z(t)$ belongs to the transversal space $\cZ(t)$.
The tangent space 
$\cT(t)$ is the root space of $A(t)$ which corresponds to the "unstable" spectral point $\lam=0$.
The key observation is that i) the symplectic-orthogonal space  
${\cal Z}(t)$ 
 does not contain the "unstable" tangent vectors, and moreover, 
ii) ${\cal Z}(t)$ 
is {\bf invariant} under the generator $A(t)$ since
$\cT(t)$ is invariant and 
$A(t)$ is the Hamiltonian operator. 
\smallskip\\
{\bf S3. Continuous and discrete components.} 
The transversal component admits further splitting $Z(t)=z(t)+f(t)$, where
$z(t)$ and $f(t)$ belong respectively 
to the  discrete and continuous spectral spaces  $\cZ_d(t)$ and $\cZ_c(t)$ 
of the generator $A(t)$ in the 
invariant space $\cZ(t)=\cZ_d(t)+\cZ_c(t)$.
 \smallskip\\
{\bf S4. Elimination of continuous component.} 
Equation (\ref{Zt}) can be projected onto $\cZ_d(t)$ and $\cZ_c(t)$.
Then the continuous transversal component $f(t)$  can be expressed via $z(t)$
and the terms $\cO(\Vert f(t))\Vert^2$
from the projection onto $\cZ_c(t)$. Substituting this expression 
into the projection onto $\cZ_d(t)$, we obtain a nonlinear cubic  equation
for $z(t)$ which includes also `higher order terms' $\cO([\Vert f(t))\Vert+|z(t)|^2]^2)$:
see equations (3.2.1)-(3.2.4) and (3.2.9)-(3.2.10) of \cite{BS2003}.
(For relativistically-invariant Ginzburg-Landau equation similar reduction  has been 
done
in  \cite[(4.9) and (4.10)]{KopK2011b}.)
 \smallskip\\
{\bf S5. Poincar\'e normal forms and Fermi Golden Rule.}
Neglecting the higher order terms,
the equation for $z(t)$ 
reduces to the Poincar\'e  normal form
which implies the decay for $z(t)$ due to the `Fermi Golden Rule' 
\cite[(1.0.12)]{BS2003}.
 \smallskip\\
{\bf S6. Method of majorants.}
A skillful interplay between the obtained decay and the extra slow evolution 
of the soliton parameters 
{\bf S1}
provides the decay for $f(t)$  and  $z(t)$ by the method of majorants. 
This decay immediately results in the asymptotics (\ref{sdw})-(\ref{glob2u}).

\subsection{Development and applications}

In \cite{BKKS2008, KKopSt2011}, these methods and results were extended i) to the 
Schr\"odinger equation interacting with nonlinear $U(1)$-invariant 
oscillators, ii)  in \cite{ IKS2011, IKV2011}, to the system \eqref{wq3}  and 
to \eqref{ML} with zero external fields, and iii) in \cite{IKV2006, KKop2006,KKopS2011}, to similar translation-invariant systems of 
Klein--\allowbreak Gordon, Schr\"odinger and Dirac equations coupled to a~particle. 
A survey of the results 
\cite{IKV2006,IKS2011,IKV2011} can be found in \cite{Im2013}.

For example, in \cite{IKV2011} we have considered solutions to 
the system \eqref{wq3} with initial data close to the solitary manifold 
(\ref{solit}) in the weighted norm 
\begin{equation}\label{ves} 
\Vert\psi\Vert_\si^2=\int \langle x\rangle^{2\si}|\psi(x)|^2dx. 
\end{equation} 
Namely, the initial state is close to soliton (\ref{solit}) with some parameters 
$v_0,a_0$: 
\begin{equation}\label{usl} 
\begin{gathered} 
\Vert\na\psi(x,0)-\na \psi_{v_0} (x-a_0)\Vert_\si 
+ 
\Vert\psi(x,0)-\psi_{v_0}(x-a_0)\Vert_\si 
+\Vert\pi(x,0)- \pi_{v_0}(x-a_0)\Vert_\si 
\\ 
+ 
|q(0)-a_0|+ 
|\dot q(0)-v_0|\le\ve, 
\end{gathered} 
\end{equation} 
where $\si>5$ and $\ve>0$ are sufficiently small. 
Moreover, we assume the Wiener condition \eqref{W1} for $k\ne 0$, while 
\begin{equation}\label{W2} 
\pa^\alpha\hat\rho(0)=0, \quad |\alpha|\le 5; 
 \end{equation} 
this  is equivalent to 
\begin{equation}\label{W2e} 
\int x^\alpha\rho(x)\,dx=0, \quad |\alpha|\le 5. 
 \end{equation} 
 Under these conditions, the main results of \cite{IKV2011} are the following 
 asymptotics: 
 \begin{equation}\label{q} 
\dot q(t)\to v_\pm, \quad q(t)\sim v_\pm t+a_\pm, \qquad t\to\pm\infty 
\end{equation} 
 (cf.\ (\ref{dq})). Moreover, the attraction to solitons (\ref{ssol}) holds, 
where the remainders 
now decay in the weighted norm in the moving frame of the particle 
(cf. (\ref{ssolh})): 
\begin{equation}\label{ssolhw} 
\Vert\na r_\pm(q(t)+x,t)\Vert_{-\si}+\Vert r_\pm(q(t)+x,t)\Vert_{-\si}+\Vert s_\pm(q(t)+x,t)\Vert_{-\si}  \xrightarrow
[t\to\pm\infty]{} 0. 
\end{equation} 
In \cite{K2002}--\cite{Kumn2013} and \cite{BC2012}, the methods and results 
\cite{BP1993}--\cite{BS2003} were extended 
to relativistically-invariant nonlinear equations. Namely, in \cite{K2002}--\cite{Kumn2013} the asymptotics of type (\ref{sdw})
were obtained for the first time for the relativistically-invariant nonlinear 
Ginzburg--\allowbreak Landau equations, 
and in \cite{BC2012}, for relativistically-invariant nonlinear Dirac equations. 
In \cite{KKK2013}, we have constructed examples of 
Ginzburg--\allowbreak Landau type
potentials providing the spectral properties 
of the linearized dynamics
imposed in \cite{K2002}--\cite{Kumn2013}.
In \cite{KKsp2013}, we have justified the eigenfunction expansions for nonselfadjoint Hamiltonian operators
which were used in \cite{K2002}--\cite{Kumn2013}. For the justification we have developed 
a special version of M.G. Krein theory of $J$-selfadjoint operators.

\medskip

In \cite{FG2014}, the system 
of type \eqref{wq3} with the Schr\"odinger equation instead of the wave equation 
is considered as a~model of the Cherenkov radiation of a tracer particle (the system (1.9)--(1.10) of  \cite{FG2014}). 
The main result of \cite{FG2014} is the long time convergence to a~soliton with a subsonic speed 
for initial solitons with supersonic speeds.
The asymptotic stability of the solitons for similar  system has been established in \cite{KKop2006}.

\medskip

Asymptotic stability of $N$-soliton solutions to nonlinear translation-invariant  
Schr\"odinger equations was studied in \cite {MMT2002}--\cite{RSS2005} 
by developing the methods of \cite{BP1993}--\cite{BS2003}. 
 \medskip 
 
 \begin{remark}
{\rm 
The asymptotics (\ref{q})--(\ref{ssolhw}) 
mean the proximity of the trajectory to the  solitary manifold in the weighted norms
under the proximity of the corresponding initial state (\ref{usl}) and 
under the Wiener condition (\ref{W1}).
The Wiener condition implies also the global 
attraction to solitons (\ref{dq})--(\ref{ssolh}). This could suggest an impression that
the Wiener condition provides
the proximity 
to the solitary manifold (\ref{usl}) 
for large times. 
 However, this impression  is erroneous since the decay (\ref{ssolh}) 
 implies the proximity  in the local energy seminorms
 which is weaker than 
 the proximity  in the weighted norms
 (\ref{usl}).}
 
\end{remark}


 \setcounter{equation}{0} 
\section{Numerical simulation of soliton asymptotics} 
 Here we describe the results of our  joint work with Arkady Vinnichenko (1945--2009) on numerical simulation of the global attraction 
 to solitons \eqref{att} and \eqref{at14},
and adiabatic effective soliton-type dynamics (\ref{effd2}) for the 
 relativistically-invariant  one-dimensional 
nonlinear wave equations \cite{KMV2004}.

\subsection{Kinks of relativistically-invariant Ginzburg--\allowbreak Landau equation} 
We have considered real solutions to 
the relativistically-invariant 1D Ginzburg--\allowbreak Landau equation, which is 
the nonlinear Klein--\allowbreak Gordon equation with polynomial nonlinearity 
\begin{equation}\label{GL} 
\ddot\psi(x,t)=\psi''(x,t)+F(\psi(x,t)),\qquad \mbox{where}\quad F(\psi):=-\psi^3+\psi. 
\end{equation} 
Since $F(\psi)=0$ for $\psi=0,\pm1$, there are three equilibrium positions 
 $S(x)\equiv 0,+1,-1$. 
 
The corresponding potential reads 
 $U(\psi)=\frac{\psi^4}{4}-\frac{\psi^2}{2}$. 
 This potential has minimum at~$\pm 1$ and maximum at~0, so the two equilibria are stable, and one is unstable. 
Such potentials with two wells are called the Ginzburg--\allowbreak Landau potentials. 
 
Besides constant stationary solutions 
 $S(x)\equiv 0,+1,-1$, 
 there is still a~non-constant steady-state "kink" solution 
 $S(x)=\tanh\frac{x}{\sqrt2}$. Its shifts and reflections 
 $\pm S(x-a) $ are also stationary solutions, as well as 
 their Lorentz transformations 
  $\pm S(\ga(x-a-vt))$ with 
 $\ga=\frac 1{\sqrt{1-v^2}}$ for
  $|v|<1$. 
  These are uniformly moving waves (i.e., solitons). 
 When the velocity $v$ is close to $\pm 1$, this kink is very compressed. 
 
\begin{figure}[htbp] 
\begin{center} 
\includegraphics[width=0.95\columnwidth]{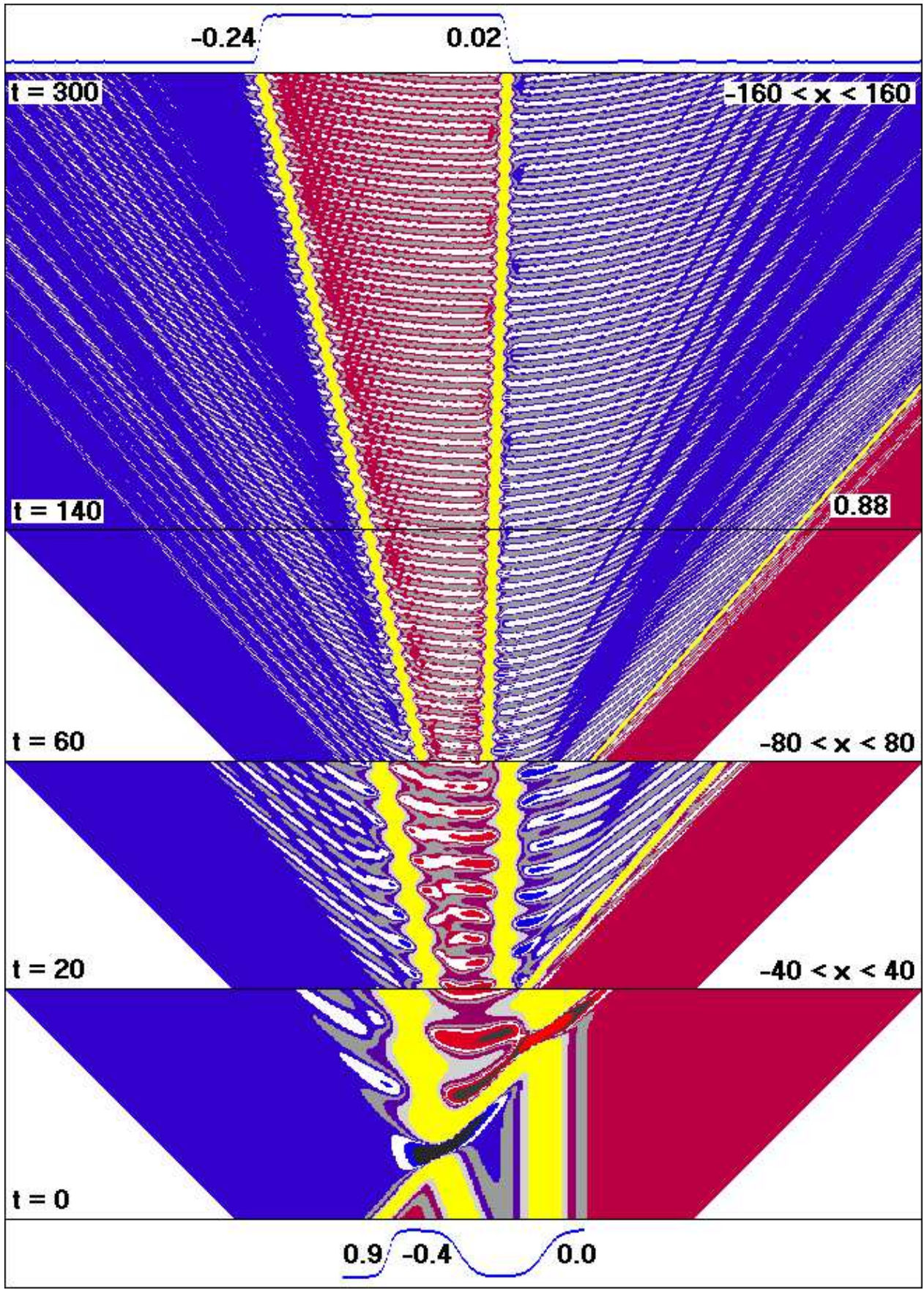} 
\caption{Decay to three kinks} 
\label{fig-4} 
\end{center} 
\end{figure}

 Equation (\ref{GL}) is 
 equivalent to the Hamiltonian system of form (\ref{w1ham}) with the Hamilton functional 
 \begin{equation}\label{hamGL} 
\cH(\psi,\pi)=\int[\frac12|\pi(x)|^2+\frac12|\psi'(x)|^2+U(\psi(x))]\,dx 
\end{equation} 
defined on the Hilbert phase space $\cE$ of states $(\psi, \pi)$ with the norm (\ref{cE}), for which 
$$ 
\psi(x)\xrightarrow
[|x|\to\infty]{}\pm 1. 
$$ 
Our numerical experiments show 
the decay of finite energy solutions to a finite collection of 
kinks and a dispersion wave that confirms the asymptotics \eqref{at14}.
One of the simulations is shown
on Fig.~\ref{fig-4}: the considered 
finite energy solution to equation (\ref{GL}) decays to three kinks. 
Here, the vertical line is the time axis and the horizontal line is the space axis. 
The spatial scale redoubles at $t=20$ and $t=60$. 
 
The red color corresponds to values $\psi> 1- \ve$, the blue one, to values $\psi <-1 + \ve$, 
and the yellow one, to values $-1+\ve <\psi < 1 + \ve$. 
Thus, the yellow stripes represents the kinks, while 
the blue and red zones outside the yellow stripes are filled with the dispersion waves 
$W(t) \Phi_+ $. 
 
At $t=0$ the solution starts from a~fairly chaotic behavior when there are no kinks. 
After 20 seconds, there are three distinct kinks, which further move almost uniformly. 
 
The left kink moves to the left with small velocity 
$v_1\approx 0.24$, the central kink is almost standing with the velocity $v_2\approx0.02$, 
and the right kink is very fast with velocity $v_3\approx0.88$. The Lorentz contraction $\sqrt{1-v_k^2}$ is clearly visible 
on this picture: 
the central kink is wide, the left one is slightly narrower, and the right one is quite narrow. 
 
Furthermore, the Einstein time delay here is also very pronounced. Namely,  
all three kinks oscillate due to presence of a~nonzero eigenvalue in the linearized equation on the kink: 
substituting $\psi (x, t)=S (x) +\ve \vp (x,t)$ into (\ref{GL}) we obtain 
$$ 
\ddot\vp(x,t)=\vp''(x,t)-2\vp(x,t)-V(x)\vp(x,t) 
$$ 
in the first order the linearized equation, where the potential 
$$ 
V(x)=3S^2(x)-3=-\frac{3}{\cosh^2\frac{x}{\sqrt2}} 
$$ 
exponentially decays for large $|x|$. 
It is a~great joy that for this potential the spectrum of the corresponding 
\textit{Schr\"odinger operator} $H:=-\frac{d^2}{dx^2}+2+V(x)$ is well known~\cite{Lamb80}. 
 Namely, the operator $H$ is non-negative, and its continuous spectrum coincides with $[2, \infty)$. 
It turns out that $H$ 
still has a~two-point discrete spectrum: the points $\lambda = 0$ and $\lambda=\frac32$. 
These pulsation, which we observe for the central slow kink, have frequency 
$\om_1\approx\sqrt{\frac32}$ and period 
 $T_1\approx2\pi/\sqrt{\frac32}\approx 5$\,s.  On the other hand, for the fast kink the ripples are much slower; i.e., the corresponding period is larger. This time delay 
 agrees with 
the Lorentz formulas.

 These agreements 
confirm the relevance of our numerical simulations of the solitons. 
Moreover, an analysis of the dispersion waves gives additional confirmations. Namely, 
the space outside the kinks in Fig.~\ref{fig-4} 
is filled with dispersion waves, whose values are very close to $\pm 1$, with the accuracy $0.01$. 
The waves satisfy, with high accuracy, the linear Klein--\allowbreak Gordon equation, which is obtained by linearization 
of the 
Ginzburg--\allowbreak Landau equation (\ref{GL}) on the stationary solutions $\psi=\pm 1$: 
$$ 
\ddot\vp(x,t)=\vp''(x,t)+2\vp(x,t). 
$$ 
The corresponding dispersion relation $\om^2 =k^2+2$ defines the group velocities of the 
wave packets, 
\begin{equation}\label{ev} 
\na\om=\fr k{\sqrt{k^2 +2}}=\pm\fr {\sqrt{\om^2 -2}}\om 
\end{equation} 
which are clearly seen in Fig.~\ref{fig-4} as straight lines  
whose propagation velocities 
approach $\pm 1$. 
This approach is explained by the limit  $|\na\om|\to 1$ for high frequencies $\om=\pm n\om_1\to\infty$ generated by the polynomial nonlinearity in (\ref{GL}). 
\begin{remark}\label{rrm}
{\rm 
These observations agree completely with the radiation mechanism summarized in Remark \ref{phys}.}
 
 \end{remark}
The nonlinearity in (\ref{GL}) is chosen so as to have well-known spectrum of the linearized equation. In the numerical experiments \cite{KMV2004} we have 
considered more general nonlinearities, and the results were qualitatively the same: 
for `any' initial data the solution again splits into a~sum of solitons. 
Numerically, this can be clearly visible, but the rigorous justification is still the matter for the future. 
 
\subsection{Numerical observation of soliton asymptotics} 
Besides the kinks our numerical experiments \cite{KMV2004} have also resulted in 
the soliton-type asymptotics (\ref{at14}) and adiabatic effective dynamics of type (\ref{effd2}) 
 for complex solutions to the 1D relativistically-invariant nonlinear wave equations 
 (\ref{NWEn}). 
Namely, we have considered the polynomial potentials of the form 
\begin{equation}\label{pop2} 
U(\psi)=a|\psi|^{2m}-b|\psi|^{2n}, 
\end{equation} 
where $a,b>0$ and $m>n=2,3,\dotsc$. 
Respectively, 
\begin{equation}\label{pop2f} 
F(\psi)=2am|\psi|^{2m-2}\psi-2bn|\psi|^{2n-2}\psi. 
\end{equation} 
The parameters $a,b,m,n$ were taken as follows: 
$$ 
\begin{array}{rrrlr} 
N~~~~~&a~~~~~&~~~~m&~~~~~~b&~~~~n\\ 
1~~~~~&1~~~~&~~~~3&~~~~~~0.61&~~~~2\\ 
2~~~~~&10~~~~&~~~~4&~~~~~~2.1&~~~~2\\ 
3~~~~~&10~~~~&~~~~6&~~~~~~8.75&~~~~5 
\end{array} 
$$ 
We have 
considered various `smooth' initial functions 
$ \psi (x, 0), \dot\psi (x, 0) $ with the support on the interval $[-20,20]$. 
The second order 
finite-difference scheme with $ \De x, \De t \sim 0.01, 0.001$ was employed. In all cases we have observed the asymptotics of type (\ref{at14}) 
with the numbers of solitons $0,1,3$ for $t> 100$.

\subsection{Adiabatic effective dynamics of relativistic solitons in external potential} 
In the numerical experiments \cite{KMV2004} was also observed 
the adiabatic effective dynamics of type (\ref{effd2}) for soliton-like solutions 
for the 1D equations 
(\ref{NWEn}) with a~slowly varying external potential (\ref{asolV}): 
\begin{equation}\label{EQp} 
\ddot \psi(x,t)=\psi''(x,t)-\psi(x,t) 
+F(\psi(x,t))-V(x)\psi(x,t), \qquad x\in\mathbb R. 
\end{equation} 
This equation is equivalent to the Hamilton system (\ref{w1ham}) with the Hamilton functional 
\begin{equation}\label{HAMp} 
\cH_V(\psi,\pi)=\int [\frac12 |\pi(x)|^2+ \frac12 |\psi^\pr (x)|^2 
+U(\psi(x))+\fr12 V(x) |\psi(x)|^2 ]\,dx.
\end{equation} 
In notations  (\ref{swen}),
the soliton-like solutions are of the form 
(cf. (\ref{asol})) 
\begin{equation}\label{swei} 
\psi(x,t)\approx e^{i\Theta(t)}\phi_{\om(t)}(\gamma_{v(t)}(x-q(t)) ).
\end{equation} 
Below we describe our numerical experiments, which qualitatively confirm 
the adiabatic effective Hamilton type dynamics
for the parameters $\Theta,\om,q$, and $v$, but its rigorous justification is still not established. 
 
\begin{figure}[htbp] 
\begin{center} 
\includegraphics[width=0.9\columnwidth]{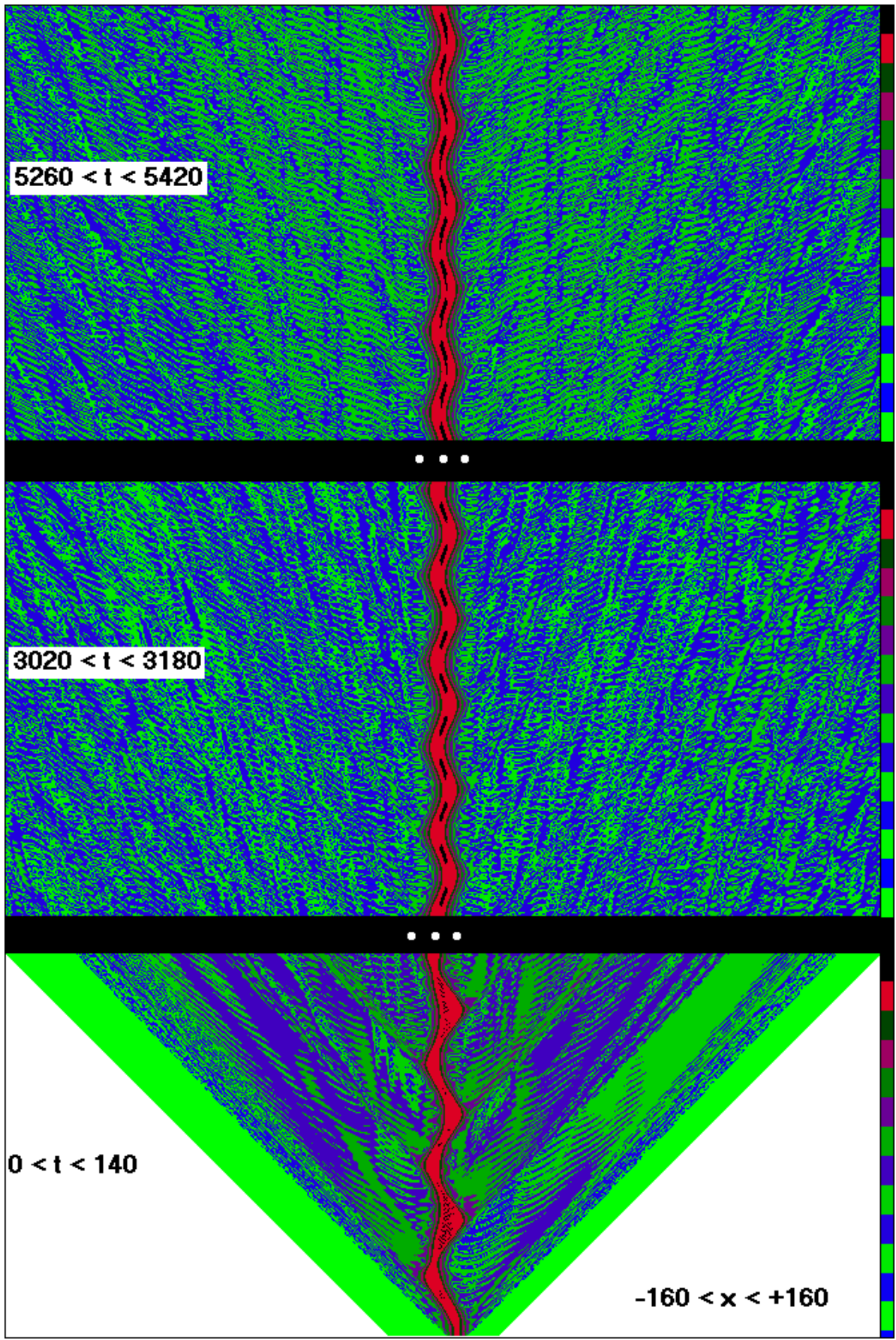} 
\caption{Adiabatic effective dynamics of solitons} 
\label{fig-7} 
\end{center} 
\end{figure}

Figure~\ref{fig-7} represents 
 a~solution to equation (\ref{EQp}) with the potential (\ref{pop2}), where $a=10$, $m=6$ 
 and $b=8.75$, $n = 5$. 
We choose 
$V(x)=-0.2\cos (0.31 x)$ 
and the initial conditions 
\begin{equation}\label{inco} 
\psi(x,0)= \phi_{\om_0}(\ga_{v_0}(x-q_0)), \qquad \dot \psi(x,0)=0, 
\end{equation} 
where $v_0=0$, $\om_0=0.6$ and $q_0=5.0$. 
Note that the initial state does not belong to the solitary manifold. An 
effective width (half-amplitude) of the solitons is in the range $[4.4, 5.6 ]$. 
It is quite small when compared with the spatial period of the potential $2\pi/0.31 \sim 20$, 
which is confirmed by numerical simulations shown on Figure~\ref{fig-7}. Namely, 
\medskip\\ 
$\bullet$ Blue and green colors represent the 
dispersion wave with 
values $|\psi(x,t)| <0.01$, while the 
red color represents the soliton with values $|\psi(x,t)|\in [0.4, 0.8]$. 
\medskip\\ 
$\bullet$ The soliton trajectory (`red snake') corresponds to 
oscillations of a~classical particle in the potential $V (x)$. 
\medskip\\ 
$\bullet$ For $0<t<140$ the solution is rather distant from the solitary manifold, and 
the radiation is intense. 
\medskip\\ 
$\bullet$ For $3020 <t< 3180 $ the solution approaches the solitary manifold, and the radiation 
weakens. The oscillation amplitude of the soliton is almost unchanged for a~long time, confirming  a Hamilton type dynamics. 
\medskip\\ 
$\bullet$ However, for $5260 <t<5420$ the amplitude of the soliton 
oscillation is halved. 
This suggests that at a~large time scale the deviation 
from the Hamilton  effective dynamics  becomes essential. 
Consequently, the effective dynamics gives a~good 
approximation only on the adiabatic time scale $t\sim \ve^{-1}$. 
\medskip\\ 
$\bullet$ The deviation from the Hamilton dynamics is due to radiation, which plays the role of dissipation. 
\medskip\\ 
$\bullet$ 
The radiation is realized as the dispersion waves which bring the energy to the infinity.
The  dispersion waves combine into uniformly moving bunches 
with discrete set of group velocities, as in Fig.~\ref{fig-4}. 
The magnitude of solutions is of order $\sim 1$ 
on the trajectory of the soliton, while the values of the dispersion waves 
is less than $0.01 $ for $t>200$, so that their energy density does not exceed $0.0001$. 
 The amplitude of the dispersion waves decays for large times.
 \medskip\\ 
$\bullet$ In the limit $t\to\pm\infty$  the soliton should converge to a static position corresponding to a local minimum 
of the potential. However, the numerical observation of this "ultimate stage" is hopeless since the rate of the convergence 
decays with the decay of the radiation.

 

\setcounter{section}{0}
\setcounter{equation}{0}
\protect\renewcommand{\thesection}{\Alph{section}}
\protect\renewcommand{\theequation}{\thesection.\arabic{equation}}
\protect\renewcommand{\thesubsection}{\thesection.\arabic{subsection}}
\protect\renewcommand{\thetheorem}{\Alph{section}.\arabic{theorem}}

\section{Attractors and quantum postulates}

The foregoing results on attractors of the nonlinear Hamilton equations were suggested by 
fundamental postulates of quantum theory, 
primarily Bohr's postulate on transitions 
between quantum stationary orbits. Namely, in 1913 Bohr suggested `Columbus's' solution 
of the problem of 
stability of atoms and molecules 
\cite{Bohr1913}, 
postulating that 
\begin{gather} 
{\parbox{153mm}{\it Atoms and molecules are permanently on  some stationary orbits $|E_m \rangle$ 
 with energies $E_m$, 
and sometimes make transitions between the orbits,}}\nonumber\bigskip\\ 
|E_m\rangle\mapsto |E_n\rangle.\label{B} 
\end{gather}  
The simplest dynamic interpretation of this postulate is the attraction to stationary orbits (\ref{atU}) 
for
any finite energy quantum trajectory $\psi(t)$. 
This means that the stationary orbits form 
a global attractor of the corresponding quantum dynamics.

However, this convergence contradicts 
the Schr\"odinger's linear equation
due to the superposition principle. 
Thus, Bohr's transitions (\ref{B}) in the linear theory do not exist. 
\medskip 

It is natural to suggest that the attraction (\ref{atU}) holds  for a~nonlinear modification of the linear Schr\"odinger theory. Namely 
it turns out that the original Schr\"odinger theory is nonlinear, because it involves interaction with the Maxwell field. 
The corresponding nonlinear Maxwell--Schr\"odinger system is contained essentially in the first Schr\"odinger's article of 1926: 
\begin{equation} \label{SM} 
\left\{ 
\begin{aligned} 
& i\dot\psi(x,t)=\ds\fr12[-i\nabla +\bA(x,t)+\bA^{\rm ext}(x,t)]^2\psi(x,t)+[A_0(x,t)+A_0^{\rm ext}(x)]\psi(x,t) 
\\ 
& \Box A_{\alpha}(x,t)=4\pi J_{\alpha}(x,t),\qquad \alpha=0,1,2,3 
\end{aligned}\right| \quad x\in\mathbb R^3, 
\end{equation} 
where the units are chosen so that $\hbar=e=m=c=1$. 
Maxwell's equations are written here in the 4-dimensional form, 
where 
$A=(A_0,\bA)=(A_0,A_1, A_2, A_3)$ 
denotes the 4-dimensional potential of the Maxwell field 
with the Lorentz  gauge $\dot A_0+\na\cdot\bA=0$, $A^{\rm ext} = (A_0^{\rm ext}, \bA^{\rm ext})$ is an external 4-potential, and 
$J=(\rho,j_1, j_2 , j_3)$ is the 4-dimensional current. 
To make these equations a~closed system, we 
must also express the density of charges and currents via the 
wave function: 
\begin{equation} \label{roj} 
J_0(x,t)=|\psi(x,t)|^2; \qquad 
J_k(x,t)=[(-i\nabla_k+A_k(x,t)+A_k^{\rm ext}(x,t))\psi(x,t)]\cdot\psi(x,t),\quad k=1,2,3; 
\end{equation} 
here `$\,\cdot\,$' denotes the scalar product of two-dimensional real vectors corresponding to complex numbers. 
In particular, 
these expressions satisfy the 
continuity equation $\dot\rho+\div j=0$ for any 
solution of the Schr\"odinger equation with arbitrary potentials \cite[Section 3.4]{K2013}. 
\medskip 

System (\ref{SM}) is non-linear in $(\psi, A)$ although the Schr\"odinger equation is formally linear 
in $\psi$. Now the question arises: 
 what should be the stationary orbits for the nonlinear hyperbolic system (\ref{SM})? 
 It is natural to suggest that these are the solutions of type 
\begin{equation}\label{ss} 
(\psi(x)e^{-i\omega t},~A(x)). 
\end{equation} 
Indeed, 
such functions give stationary distributions of charges and currents (\ref{roj}). 
 Moreover, these functions are the trajectories of one-parameter subgroups of the 
 symmetry group $U(1)$ of the system (\ref{SM}). Namely, for any solution 
 $(\psi(x,t), A(x,t))$ and $\theta\in\R$ the functions 
\begin{equation}\label{Ut}
U_\theta(\psi(x,t),~A(x,t)):=(\psi(x,t)e^{i\theta},~A(x,t))
\end{equation} 
are also solutions. 
The same remarks apply to the Maxwell--\allowbreak Dirac system introduced by Dirac in 1927: 
\begin{equation}\label{DM} 
\left\{ 
 \begin{aligned} 
& \sum_{\al=0}^3 \ga^\al[i\ds \na_\al - 
A_\al(x,t)- 
A_\al^{\rm ext}(x,t)]\psi(x,t) = 
m \psi(x,t)\\ 
& \Box\, A_\al(x,t) =J_\al(x,t) := \ov{\psi(x,t)}\ga^0\ga_\al\psi(x,t), \quad \al=0,\dotsc,3\ \ 
 \end{aligned}\right| \quad x\in\mathbb R^3, 
\end{equation} 
where $\na_0:=\pa_t$. 
Thus, Bohr's transitions (\ref{B}) for the systems (\ref{SM}) and (\ref{DM}) 
 with a static external potential $A^{\rm ext}(x,t)=A^{\rm ext}(x)$ 
 can be interpreted as the long-time asymptotics
 \begin{equation}\label{as} 
(\psi(x,t),~A(x,t))\sim (\psi_\pm(x)e^{-i\omega_\pm t},~A_\pm(x,t)), \qquad t\to\pm\infty 
\end{equation} 
for every finite energy solution, where the asymptotics hold in a local norm.
Obviously, the maps $U_\theta$ form the group isomorphic to $U(1)$, and 
the functions (\ref{ss}) are the trajectories of its one-parametric subgroups.
Hence, the asymptotics (\ref{as}) correspond to our general conjecture 
(\ref{at10})
with the symmetry group $U(1)$.
\medskip

 Furthermore, in the case of zero external potentials   
these systems  are 
translation-invariant.  
Respectively, for their solutions one should expect the soliton asymptotics of type \eqref{at14}
\begin{equation}\label{SA} 
(\psi(x,t), A(x,t))\sim 
\ds\sum\limits_{k} (\psi_\pm^k(x-v^k_\pm t) 
e^{i\Phi_\pm^k(x,t)},~A_\pm^k(x-v^k_\pm t))+(\varphi_\pm(x,t),~A_\pm(x,t)), \qquad t\to\pm\infty, 
\end{equation} 
where the asymptotics hold in a global norm.
Here $\Phi_\pm^k(x,t) $ are suitable phase functions, and each term-soliton is a~solution to the  
corresponding nonlinear system, while $\varphi_\pm(x,t)$ and $A_\pm(x,t)$ 
represent  some  dispersion waves which are 
solutions to the free 
Schr\"odinger and Maxwell equations respectively. 
The existence of the solitons for the Maxwell-Dirac system is established
 in \cite{EGS}. 
 \medskip
 
The asymptotics (\ref{as}) and (\ref{SA}) are not proved yet for the 
Maxwell-Schr\"odinger and
Maxwell-Dirac equations 
(\ref{SM}) and (\ref{DM}). 
One could expect that 
these asymptotics
should follow by suitable modification of the arguments from Section 3.  
Namely, let the time spectrum
of an omega-limit trajectory 
$\psi(x,t)$ contain at least two different frequencies $\om_1\ne\om_2$:
for example, $\psi(x,t)=\psi_1(x)e^{-i\om_1 t}+\psi_2(x)e^{-i\om_2 t}$.
Then the currents $J_\al(x,t)$ in the systems (\ref{SM}) and (\ref{DM})
contains the terms with the harmonics  $e^{-i\De t}$ and 
 $e^{i\De t}$, where $\De:=\om_1-\om_2\ne 0$. 
Thus the nonlinearity inflates the spectrum as in $U(1)$-invariant 
equations, considered in Section 3.

Further, these 
time-dependent harmonics on the right hand side of the Maxwell equations 
induce the radiation of an electromagnetic wave
with the frequency $\Delta$
according to  the limiting amplitude  principle (\ref{lap})
since the continuous spectrum of the Maxwell generator is the whole line 
$\R$. 
Finally, this radiation brings the energy to infinity which is impossible for omega-limit trajectories. 
This contradiction suggests the validity of  
the one-frequency asymptotics (\ref{as}).
 Let us note that
the spectrum of the radiation
contains the  difference
$\om_1-\om_2$ in accordance with the second Bohr postulate. 

We have justified
 similar arguments rigorously  for  $U(1)$-invariant 
equations (\ref{KG1}) and (\ref{KGN})--(\ref{Dn}). 
However, for the systems (\ref{SM}) and (\ref{DM})
the rigorous justification is still an open problem.

\end{document}